\documentclass[twocolumn,superscriptaddress,floatfix,aps,prd,nofootinbib,showpacs]{revtex4}

\usepackage{graphicx}

\begin{document}

\title{Quasiequilibrium black hole-neutron star
binaries in general relativity}

\author{Keisuke Taniguchi}
\affiliation{Department of Physics, University of Illinois at
  Urbana-Champaign, Urbana, Illinois 61801, USA}
\author{Thomas W. Baumgarte}
\altaffiliation{Also at: Department of Physics, University of Illinois
  at Urbana-Champaign, Urbana, Illinois 61801, USA}
\affiliation{Department of Physics and Astronomy, Bowdoin College,
  Brunswick, Maine 04011, USA}
\author{Joshua A. Faber}
\affiliation{Department of Physics, University of Illinois at
  Urbana-Champaign, Urbana, Illinois 61801, USA}
\author{Stuart L. Shapiro}
\altaffiliation{Also at: Department of Astronomy and NCSA, University
  of Illinois at Urbana-Champaign, Urbana, Illinois 61801, USA}
\affiliation{Department of Physics, University of Illinois at
  Urbana-Champaign, Urbana, Illinois 61801, USA}

\date{October 24, 2007}

\begin{abstract}

We construct quasiequilibrium sequences of black hole-neutron star
binaries in general relativity.  We solve Einstein's constraint
equations in the conformal thin-sandwich formalism, subject to black
hole boundary conditions imposed on the surface of an excised sphere,
together with the relativistic equations of hydrostatic equilibrium.
In contrast to our previous calculations we adopt a flat spatial
background geometry and do not assume extreme mass ratios.  We adopt a
$\Gamma=2$ polytropic equation of state and focus on irrotational
neutron star configurations as well as approximately nonspinning black
holes.  We present numerical results for ratios of the black hole's
irreducible mass to the neutron star's ADM mass in isolation of
$M_{\rm irr}^{\rm BH}/M_{\rm ADM,0}^{\rm NS} =1$, 2, 3, 5, and 10.  We
consider neutron stars of baryon rest mass $M_{\rm B}^{\rm NS}/M_{\rm
B}^{\rm max} =$ 83\% and 56\%, where $M_{\rm B}^{\rm max}$ is the
maximum allowed rest mass of a spherical star in isolation for our
equation of state. For these sequences, we locate the onset of tidal
disruption and, in cases with sufficiently large mass ratios and
neutron star compactions, the innermost stable circular orbit.  We
compare with previous results for black hole-neutron star binaries and
find excellent agreement with third-order post-Newtonian results,
especially for large binary separations. We also use our results to
estimate the energy spectrum of the outgoing gravitational radiation
emitted during the inspiral phase for these binaries.
\end{abstract}


\maketitle

\section{Introduction}

Coalescing compact binaries composed of neutron stars and/or black
holes are among the most promising sources of gravitational waves for
ground-based laser interferometers \cite{LIGO,GEO,TAMA,VIRGO}.
Observations of short gamma-ray bursts (SGRBs) by the {\it Swift} and
{\it HETE-2} satellites (see, e.g.~\cite{Barge06} and references
therein) suggest that their central engine may well be the merger remnant 
of compact binaries that contain a neutron star, namely the remnants of
black hole-neutron star (BHNS) binaries or binary neutron star (BNS)
systems \cite{ShibaT06,FaberBST06,PriceR06,OechsJ06}.  Motivated by
these considerations, significant theoretical effort has gone into the
modeling of these binaries.

So far, most studies of BHNS binaries have been performed within the
framework of Newtonian gravity in either some or all aspects of the
calculation (see, e.g.~
\cite{Chand69,Fishb73,LaiRS93,LaiW96,TanigN96,Shiba96,UryuE99,WiggiL00,IshiiSM05}
for quasiequilibrium calculations and
\cite{Mashh75,CarteL83,Marck83,LeeK99,Lee00,RosswSW04,KobayLPM04} for
dynamical simulations).  More recently, several groups have initiated
studies of BHNS binaries in a fully relativistic framework, both for
quasiequilibrium models
\cite{Mille01,BaumgSS04,TanigBFS05,TanigBFS06,Grand06} and dynamical
simulations
\cite{FaberBSTR06,FaberBST06,SopueSL06,LofflRA06,ShibaU06}.

In our own effort, we initially studied BHNS quasiequilibrium models
under the assumption of extreme mass ratios, where the mass of the
black hole is much greater than that of the neutron star
\cite{BaumgSS04,TanigBFS05,FaberBSTR06,FaberBST06}.  This assumption
is very appealing computationally and therefore a natural first step
in a systematic study of these binaries, but breaks down for the
astrophysically more interesting comparable-mass binaries.  For the
latter, the neutron star is tidally disrupted outside of the black
hole's innermost stable circular orbit (ISCO) (see, e.g.~
\cite{IshiiSM05}) and may form an accretion disk, which is
considered crucial for launching of a gamma-ray burst.  The
observation of such a process would provide a wealth of astrophysical
information \cite{Valli00}.  In a subsequent study of quasiequilibrium
BHNS models \cite{TanigBFS06}, we therefore relaxed the assumption of
extreme mass ratios and studied comparable-mass binaries (see also
\cite{Grand06,ShibaU06}).

We construct quasiequilibrium models by solving
Einstein's constraint equations in the conformal thin-sandwich
formalism (see, e.g.~\cite{York99,Cook00,BaumgS03}).  These equations
constrain only some of the gravitational fields; others, namely the
conformal background solution, are freely specifiable and have to be
chosen before the equations can be solved (see Section \ref{Sec:IIA}
below).  In our previous studies
\cite{BaumgSS04,TanigBFS05,TanigBFS06} we chose a background
solution describing a single black hole in Kerr-Schild
coordinates (in \cite{TanigBFS05} we also include results for a 
spatially flat
background for extreme mass ratio binaries).
This seemed appealing in a number of ways; for example, it allows for
black hole spin and the coordinate system extends smoothly into the
black hole interior. However, in \cite{TanigBFS06} we found sizable
deviations from post-Newtonian (PN) results which we could attribute,
at least in part, to the choice of the background solution
\cite{TanigBFS05,Grand06}.  Moreover, the
recent black hole boundary conditions of Cook and Pfeiffer
\cite{CookP04} (see also \cite{CaudiCGP06}) allow for black hole
rotation and penetrating coordinates even for a spatially flat background
solution, which simplifies the equations significantly.

In this paper we therefore revisit quasiequilibrium models of BHNS
binaries.  We now use a spatially flat background solution instead of a
Kerr-Schild metric (compare \cite{Grand06}), and we also adopt a new
decomposition of the field variables which significantly reduces the
numerical error that was still present in \cite{TanigBFS06} (see
Appendix \ref{appendix:eq}).  We construct sequences of BHNS binaries
in quasicircular orbits for fixed black hole irreducible masses and
neutron star baryon rest masses, and focus on irrotational
neutron stars orbiting approximately nonspinning black holes (we
will implement a more accurate condition for nonspinning black holes
in the future, following \cite{CaudiCGP06}).  We find excellent
agreement with post-Newtonian results for large binary separations.  We
track cusp formation to locate the onset of tidal disruption, and, for
large mass ratios and neutron star compactions, find turning-points on a
binding energy curve to locate the ISCO.  For those sequences that
encounter an ISCO we find reasonable agreement with the ISCO of test
particles around Schwarzschild black holes.

The paper is organized as follows. We briefly review the basic equations
in Section II.  We present numerical results in Section III, and
provide a comparison with the results of
\cite{TanigBFS05,TanigBFS06,Grand06} in Section IV.
In Section V we briefly summarize our findings.

Throughout this paper we adopt geometric units with $G=c=1$, where $G$
denotes the gravitational constant and $c$ the speed of light. Latin
and Greek indices denote purely spatial and spacetime components,
respectively.

\section{Formulation}

In this Section we briefly review the equations we solve to obtain
a BHNS binary in quasiequilibrium. For a more detailed discussion we
refer to the review articles \cite{Cook00,BaumgS03} as well as Sec.~II
of \cite{GourgGTMB01} for the hydrostatics.

\subsection{Gravitational field equations}
\label{Sec:IIA}

The line element in $3+1$ form is written as
\begin{eqnarray}
  ds^2 &=& g_{\mu \nu} dx^{\mu} dx^{\nu} \nonumber \\
  &=& -\alpha^2 dt^2 + \gamma_{ij} (dx^i +\beta^i dt)
  (dx^j +\beta^j dt),
\end{eqnarray}
where $g_{\mu \nu}$ is the spacetime metric, $\alpha$ the lapse
function, $\beta^i$ the shift vector, and $\gamma_{ij}$ the spatial
metric induced on a spatial slice $\Sigma$.  The spatial metric
$\gamma_{ij}$ is further decomposed according to $\gamma_{ij} = \psi^4
\tilde{\gamma}_{ij}$, where $\psi$ denotes the conformal factor and
$\tilde{\gamma}_{ij}$ the background spatial metric. We also decompose
the extrinsic curvature $K^{ij}$ into a trace $K$ and a traceless part
$\tilde{A}^{ij}$ according to
\begin{equation}
  K^{ij} = \psi^{-10} \tilde{A}^{ij} +{1 \over 3} \gamma^{ij} K.
\end{equation}
The Hamiltonian constraint then becomes
\begin{equation} \label{eq:ham}
  \tilde{\nabla}^2 \psi = -2\pi \psi^5 \rho +{1 \over 8} \psi
  \tilde{R} +{1 \over 12} \psi^5 K^2 - {1 \over 8} \psi^{-7}
  \tilde{A}_{ij} \tilde{A}^{ij}. 
\end{equation}
Here $\tilde{\nabla}^2 = \tilde \gamma^{ij} \tilde \nabla_i \tilde
\nabla_j$ is the covariant Laplace operator, $\tilde \nabla_i$ the
covariant derivative, $\tilde{R}_{ij}$ the Ricci tensor, and
$\tilde{R}=\tilde{\gamma}^{ij} \tilde{R}_{ij}$ the scalar curvature,
all associated with the conformal background metric
$\tilde{\gamma}_{ij}$.

We employ the conformal thin-sandwich decomposition of the Einstein
equations \cite{York99}.  In this decomposition, we use the
evolution equation for the spatial metric to express the
traceless part of the extrinsic curvature in terms of the time
derivative of the background metric, $\tilde{u}_{ij}\equiv\partial_t
\tilde{\gamma}_{ij}$, and the gradients of the shift vector.
Under the assumption of
equilibrium, i.e., $\tilde{u}_{ij}=0$ in a corotating coordinate
system, the traceless part of the extrinsic curvature reduces to
\begin{equation}
  \tilde{A}^{ij} ={\psi^6 \over 2\alpha} \Bigl( \tilde{\nabla}^i
  \beta^j +\tilde{\nabla}^j \beta^i -{2 \over 3} \tilde{\gamma}^{ij}
  \tilde{\nabla}_k \beta^k \Bigr). \label{eq:taij}
\end{equation}
Inserting Eq.~(\ref{eq:taij}) into the momentum constraint we
obtain
\begin{eqnarray} \label{eq:mom}
  &&\tilde{\nabla}^2 \beta^i +{1 \over 3} \tilde{\nabla}^i
  (\tilde{\nabla}_j \beta^j) +\tilde{R}^i_j \beta^j \nonumber \\
  &&= 16\pi \alpha \psi^4 j^i
  + 2\tilde{A}^{ij} \tilde{\nabla}_j (\alpha \psi^{-6})
  +{4 \over 3} \alpha \tilde{\gamma}^{ij} \tilde{\nabla}_j K.
\end{eqnarray}
For the construction of quasiequilibrium data it is also reasonable to
assume $\partial_t K =0$ in a corotating coordinate system.  The trace
of the evolution equation for the extrinsic curvature then yields
\begin{eqnarray} \label{eq:Kdot}
  \tilde{\nabla}^2 \alpha &=& 4\pi \alpha \psi^4 (\rho +S)
  +{1 \over 3} \alpha \psi^4 K^2 + \psi^4 \beta^i \tilde{\nabla}_i K
  \nonumber \\
  &&+ \alpha \psi^{-8} \tilde{A}_{ij} \tilde{A}^{ij}
  - 2 \tilde{\gamma}^{ij} \tilde{\nabla}_i \alpha \tilde{\nabla}_j
  \ln \psi.
\end{eqnarray}

Equations (\ref{eq:ham}), (\ref{eq:mom}) and (\ref{eq:Kdot}) provide
equations for the lapse function $\alpha$, the shift vector $\beta^i$,
and the
conformal factor $\psi$, while $\tilde A^{ij}$ can be found from
Eq.~(\ref{eq:taij}).  The conformally related spatial metric
$\tilde{\gamma}_{ij}$ and the trace of the extrinsic curvature $K$
remain freely specifiable, and have to be chosen before we can solve
the above equations (note that we have already set to zero the time
derivatives of these quantities, which are also freely specifiable).
In \cite{TanigBFS06} we identified these quantities with the
corresponding quantities for a Schwarzschild black hole in Kerr-Schild
coordinates.  Instead, we now assume a flat background
$\tilde{\gamma}_{ij} = \eta_{ij}$, where $\eta_{ij}$ denotes a flat
spatial metric, and maximal slicing $K=0$.  In Cartesian coordinates,
Eqs.~(\ref{eq:ham}), (\ref{eq:mom}) and (\ref{eq:Kdot}) then reduce to
\begin{eqnarray}
  &&\underline{\Delta} \psi = -2\pi \psi^5 \rho - {1 \over 8} \psi^{-7}
  \tilde{A}_{ij} \tilde{A}^{ij}, \label{eq:ham_constr} \\
  &&\underline{\Delta} \beta^i +{1 \over 3} \partial^i
  (\partial_j \beta^j) = 16\pi \alpha \psi^4 j^i
  + 2\tilde{A}^{ij} \partial_j (\alpha \psi^{-6}),
  \label{eq:mom_constr} \\
  &&\underline{\Delta} \alpha = 4\pi \alpha \psi^4 (\rho +S)
  + \alpha \psi^{-8} \tilde{A}_{ij} \tilde{A}^{ij} \nonumber \\
  &&\hspace{30pt}- 2 \eta^{ij} \partial_i \alpha \partial_j \ln \psi,
  \label{eq:trace_evolv}
\end{eqnarray}
where $\underline{\Delta}$ and $\partial_i$ denote the flat Laplace
operator and the flat partial derivative, while Eq.~(\ref{eq:taij})
becomes
\begin{equation}
  \tilde{A}^{ij} ={\psi^6 \over 2\alpha} \Bigl( \partial^i
  \beta^j +\partial^j \beta^i -{2 \over 3} \eta^{ij}
  \partial_k \beta^k \Bigr).
\end{equation}
For numerical purposes we further decompose the variables and their
equations into parts associated with the black hole and the neutron
star.  For details of this decomposition we refer to Appendix
\ref{appendix:eq}, but we note that we have found this decomposition
crucial for improving the accuracy of the calculations.

The matter terms on the right-hand side of Eqs.~(\ref{eq:ham_constr}),
(\ref{eq:mom_constr}), and (\ref{eq:trace_evolv}) are derived from the
projections of the stress-energy tensor $T_{\mu \nu}$ into the spatial
slice $\Sigma$.  Assuming an ideal fluid, we have
\begin{equation}
  T_{\mu \nu} = (\rho_0 +\rho_i +P) u_{\mu} u_{\nu} + P g_{\mu \nu},
\end{equation}
where $u_{\mu}$ is the fluid 4-velocity, $\rho_0$ the baryon rest-mass
density, $\rho_i$ the internal energy density, and $P$ the pressure.
Denoting the future-oriented unit normal to $\Sigma$ as $n_{\mu}$,
the relevant projections of  $T_{\mu \nu}$ are
\begin{eqnarray}
  &&\rho = n_{\mu} n_{\nu} T^{\mu \nu}, \\
  &&j^i = -\gamma^i_{\mu} n_{\nu} T^{\mu \nu}, \\
  &&S_{ij} = \gamma_{i \mu} \gamma_{j \nu} T^{\mu \nu}, \\
  &&S = \gamma^{ij} S_{ij}.
\end{eqnarray}

\subsection{Hydrostatic equations}

The matter in the neutron star interior has to satisfy the
relativistic equations of hydrodynamics.  For stationary
configurations, the relativistic Euler equation can be integrated once
to yield
\begin{equation}
  h \alpha {\gamma \over \gamma_0} = {\rm constant},
\end{equation}
where $h=(\rho_0 +\rho_i +P)/\rho_0$ is the fluid specific enthalpy,
and $\gamma$ and $\gamma_0$ are Lorenz factors between the fluid and
the rotating frame, and the rotating frame and the inertial frame (see
Sec. II.C. of \cite{TanigBFS05} for
the definitions).  For irrotational fluids, the fluid 3-velocity with
respect to the inertial observer, $U^i$, can be expressed in terms of
the gradient of a velocity potential $\Psi$ as
\begin{equation}
  U^i = {\psi^{-4} \over \alpha u^t h} \tilde{\nabla}^i \Psi,
\end{equation}
where $u^t$ is the time component of the fluid 4-velocity $u^{\mu}$.
Having taken into account the expression of the 3-velocity, the
equation of continuity becomes
\begin{equation}
  {\rho_0 \over h} \nabla^{\mu} \nabla_{\mu} \Psi +(\nabla^{\mu} \Psi)
  \nabla_{\mu} \Bigl( {\rho_0 \over h} \Bigr) = 0,
\end{equation}
where $\nabla_{\mu}$ denotes the covariant derivative associated with
$g_{\mu \nu}$.  We refer to \cite{GourgGTMB01} for a more detailed
derivation of the hydrostatic equations, and to
\cite{TanigG02,TanigG03,BejgeGGHTZ05} for BNS applications.

\subsection{Equation of state}

We adopt a polytropic equation of state in the form
\begin{equation}
  P =\kappa \rho_0^{\Gamma},
\end{equation}
where $\Gamma$ denotes the adiabatic index and $\kappa$ a constant.
In this paper, we focus on the case $\Gamma=2$. Using this equation of
state, the definition of the specific enthalpy
$h=(\rho_0 +\rho_i +P)/\rho_0$, and a thermodynamic relation
(Gibbs-Duhem relation)
\begin{equation}
  {d h \over h} = {dP \over \rho_0 +\rho_i +P},
\end{equation}
we obtain the internal energy density
\begin{equation}
  \rho_i ={\kappa \over \Gamma -1} \rho_0^{\Gamma}.
\end{equation}

Since dimensions enter the problem only through the constant $\kappa$,
it is convenient to rescale all dimensional quantities with respect to
the polytropic length scale
\begin{equation}
  R_{\rm poly} \equiv \kappa^{1/(2\Gamma-2)}. \label{eq:poly_units}
\end{equation}

\subsection{Boundary conditions}

In order to solve the gravitational field equations
(\ref{eq:ham_constr}), (\ref{eq:mom_constr}), and
(\ref{eq:trace_evolv}), we have to set appropriate boundary
conditions on two different boundaries: outer boundaries at
spatial infinity and inner boundaries on the black hole horizons.

The boundary conditions at spatial infinity follow from the assumption
of asymptotic flatness.  With the help of a radial coordinate
transformation $u=1/r$ in the external computational domain our
computational grid extends to spatial infinity
\cite{BonazGM98,GourgGTMB01}, and we can impose the exact boundary
conditions
\begin{eqnarray}
  &&\psi |_{r \rightarrow \infty} = 1, \\
  &&\beta^i |_{r \rightarrow \infty} = (\mbox{\boldmath $\Omega$}
  \times \mbox{\boldmath $R$})^i, \\
  &&\alpha |_{r \rightarrow \infty} = 1,
\end{eqnarray}
where $\Omega$ is the orbital angular velocity of the binary system
measured at infinity, and $\mbox{\boldmath $R$}=(X,~Y,~Z)$ is a
Cartesian coordinate which origin is located at the center of mass of
the binary system. Here, we express the
shift vector $\beta^i$ in a corotating coordinate system that we
adopt throughout our calculation.  In an inertial coordinate
system, the shift vector would tend to zero at spatial infinity,
while in the
corotating coordinate system of the numerical code the shift
vector diverges at spatial infinity.  For computational purposes it is
therefore convenient to write the shift vector as a sum of the rotational
shift term $\beta^i_{\rm rot} \equiv (\mbox{\boldmath $\Omega$} \times
\mbox{\boldmath $R$})^i$ and a residual part (which tends to zero at
spatial infinity), and solve the equations only for the latter
(see Appendix \ref{appendix:eq} for a detailed explanation of the
decomposition of the metric quantities and related equations).

The inner boundary conditions arise from the excision of the black
hole interior.  The assumption that the black hole is in equilibrium
leads to a set of boundary conditions for the conformal factor and
shift vector \cite{CookP04} (see also \cite{Cook02,CaudiCGP06} as well as
the related isolated horizon formalism, e.g.~\cite{AshteK04,GourgJ06}).
The boundary condition for the conformal factor is
\begin{equation}
  \tilde{s}^k \tilde{\nabla}_k \ln \psi \Bigl|_{\cal S}
  =-{1 \over 4} (\tilde{h}^{ij} \tilde{\nabla}_i \tilde{s}_j
  -\psi^2 J) \Bigl|_{\cal S},
\end{equation}
where $s^i \equiv \psi^{-2} \tilde{s}^i$ is the outward pointing unit
vector normal to the excision surface and $h_{ij}$ is the induced
metric on the excision surface,
$h_{ij} \equiv \psi^4 \tilde{h}_{ij} = \gamma_{ij} - s_i s_j$.  The
quantity $J$ is computed from the projection of the extrinsic
curvature $K_{ij}$ as $J \equiv h^{ij} K_{ij}$.  The boundary
condition on the normal component of the shift vector is
\begin{equation}
  \beta_{\perp} |_{\cal S} = \alpha |_{\cal S}, \label{eq:beta_perp}
\end{equation}
while the tangential components have to satisfy
\begin{equation}
  \beta_{\parallel}^i |_{\cal S} = \epsilon^{i}_{jk} \Omega_r^j x^k.
  \label{eq:beta_para}
\end{equation}
Here $\Omega_r^j$ is the black hole spin angular velocity vector,
which we take to be aligned with the $Z$-axis, and $x^k$ is a
Cartesian coordinate centered on the black hole.  To construct
approximately nonspinning black holes, we set the shift vector
according to Eqs.~(\ref{eq:beta_perp}) and (\ref{eq:beta_para}) with
$\Omega_r = \Omega$.  This assignment corresponds to a ``leading-order
approximation'' in the language of \cite{CaudiCGP06}, and we plan to
implement their more accurate condition for nonspinning black holes in
the future.  According to \cite{CookP04}, the boundary condition on
the lapse function can be chosen freely. In this paper, we choose a
Neumann boundary condition
\begin{equation} \label{eq:lapse_bound}
  {d \alpha \over dr} \Bigl|_{\cal S} =0
\end{equation}
on the excised surface, where $r$ is a radial isotropic coordinate.

\subsection{Orbital angular velocity}

In this paper we take the rotational axis of the binary system to be
the $Z$-axis, and the line connecting the black hole and neutron star
centers to be the $X$-axis.  To impose a quasicircular orbit we
require a force balance along the $X$-axis at the center of the
neutron star, which results in the condition \cite{GourgGTMB01}
\begin{equation}
  {\partial \ln h \over \partial X} \Bigl|_{(X_{\rm NS},Y_{\rm NS},0)}
  = 0.
\end{equation}
Here $X_{\rm NS}$ and $Y_{\rm NS}$ denote the $X$ and $Y$-coordinates
of the center of the neutron star relative to the rotational axis of
the binary system.  Imposing this condition determines the orbital
angular velocity $\Omega$.  We confirm that the orbital angular
velocity obtained by this method agrees with that obtained by
requiring the enthalpy at two points on the neutron star's surface to be
equal \cite{BaumgSS04} within one part in $10^{-6}$.

\subsection{Global quantities}

It is reasonable to require that the irreducible mass of the black
hole and the baryon rest mass of the neutron star are conserved during
the inspiral of the BHNS binaries.  For such a constant-mass sequence
we then monitor the Arnowitt-Deser-Misner (ADM) mass, the Komar mass,
and the total angular momentum, all of which are defined globally as
follows.

The irreducible mass of the black hole is defined as
\begin{equation}
  M_{\rm irr}^{\rm BH} \equiv \sqrt{A_{\rm EH} \over 16\pi},
\end{equation}
where $A_{\rm EH}$ is the proper area of the event horizon.  In
practice we approximate this area with that of the apparent horizon,
$A_{\rm AH}$, which is computed from an integral on the excision
surface ${\cal S}$,
\begin{equation}
  A_{\rm AH} = \int_{\cal S} \psi^4 d^2 x.
\end{equation}
The baryon rest mass of the neutron star is
\begin{equation}
  M_{\rm B}^{\rm NS} = \int \rho_0 u^t \sqrt{-g} d^3 x,
\end{equation}
where $g$ is the determinant of $g_{\mu \nu}$.  In our case we 
have $\sqrt{-g} = \alpha \psi^6$, so that 
\begin{equation}
  M_{\rm B}^{\rm NS} = \int \rho_0 \alpha u^t \psi^6 d^3 x.
\end{equation}
In the polytropic units of Eq.~(\ref{eq:poly_units}) we normalize the
baryon rest mass according to
\begin{equation}
  \bar{M}_{\rm B}^{\rm NS} \equiv {M_{\rm B}^{\rm NS} \over
    R_{\rm poly}},
\end{equation}
whereby $\bar{M}_{\rm B}^{\rm NS}$ is the rest mass of the polytrope 
for polytropic constant $\kappa = 1$.

We express the ADM mass in isotropic Cartesian coordinates as
\begin{equation}
  M_{\rm ADM} =-{1 \over 2\pi} \oint_{\infty} \partial^i \psi dS_i.
\end{equation}
The Komar mass can be written as
\begin{equation}
  M_{\rm Komar} ={1 \over 4\pi} \oint_{\infty} \partial^i \alpha dS_i,
\end{equation}
where we use the fact that the shift vector falls off sufficiently
rapidly.  The total angular momentum is
\begin{equation}
  J_i ={1 \over 16\pi} \epsilon_{ijk} \oint_{\infty} (X^j K^{kl} -X^k
  K^{jl}) dS_l,
\end{equation}
where $X^i$ is a spatial Cartesian coordinate relative to the center
of mass of the binary system.  Finally, the linear momentum is
\begin{equation}
  P^i ={1 \over 8\pi} \oint_{\infty} K^{ij} dS_j,
\end{equation}
where we have assumed maximal slicing $K=0$.

During the iteration, we require that the linear momentum vanishes.  We
enforce this by changing the position of the black hole and neutron
star relative to the location of the axis of rotation.  Stated
differently, this condition fixes the location of the axis of
rotation.

We define the binding energy of the binary system as
\begin{equation}
  E_{\rm b} = M_{\rm ADM} - M_0, \label{eq:bindene}
\end{equation}
where $M_0$ is the ADM mass of the binary system at infinite orbital
separation, as defined by the sum of the irreducible mass of the
isolated black hole and the ADM mass of an isolated neutron star with
the same baryon rest mass,
\begin{equation}
  M_0 \equiv M_{\rm irr}^{\rm BH} + M_{\rm ADM,0}^{\rm NS}.
\end{equation}

In order to measure a global error in the numerical results, we define
the virial error as the fractional difference between the ADM mass and
Komar mass,
\begin{equation}
  \delta M \equiv \Bigl| {M_{\rm ADM} - M_{\rm Komar} \over M_{\rm
  ADM}} \Bigr|. \label{eq:virial}
\end{equation}

\section{Numerical results}

Our numerical code is based on the spectral method {\sc Lorene}
library routines developed by the Meudon relativity group
\cite{Lorene}.  
Our computational grid depends on the orbital separation, the mass of
the neutron star, and the mass ratio. For example, the grid is divided
into 10 (8) domains for the black hole (neutron star) at large
separations for a neutron star mass of $\bar{M}_{\rm B}^{\rm
NS}=0.15$ and mass ratio $M_{\rm irr}^{\rm BH}/M_{\rm ADM,0}^{\rm
NS}=5$.  Each domain for the black hole is covered by $N_r \times
N_{\theta} \times N_{\phi}=41 \times 33 \times 32$ collocation points
and those for the neutron star by $25 \times 17 \times 16$, where
$N_r$, $N_{\theta}$, and $N_{\phi}$ denote the collocation points in
the radial, polar, and azimuthal directions, respectively.
We use a larger number of collocation points for the black hole
domains than the neutron star domains because the
source terms of the gravitational field equations for the black hole
vary slightly around the neutron star position. The amplitude of this
excess in the source terms is small but significantly affects the
accuracy of the results. In order to resolve this source in the black
hole domains, which are centered on the
black hole, we need a higher angular resolution. On the other hand,
the source terms of the gravitational field equations for the neutron
star have large contributions only near the neutron star, and drop off
monotonically away from the neutron star.  Thus a smaller angular
resolution is sufficient in the neutron star domains, which are centered
on the neutron star.  (See Appendix
\ref{appendix:eq} for the decomposition of the gravitational field
equations into the black hole and neutron star parts.)

As stated above, we choose $\Gamma=2$ for the adiabatic index of the
polytropic equation of state in this paper, and compute several
different constant-mass inspiral sequences.  We consider neutron stars
of rest mass $\bar{M}_{\rm B}^{\rm NS}=0.15$ and 0.10, with
corresponding compactions of
$M_{\rm ADM,0}^{\rm NS}/R_0=0.145$ and 0.0879.  Here $R_0$ is
the areal radius of the neutron star in isolation.  The maximum baryon
rest mass for spherical $\Gamma=2$ polytropes in isolation is $\bar M_{\rm
B}^{\rm max} = 0.180$.  Our more compact polytrope with $\bar M_{\rm
B}^{\rm NS}=0.15$ has a compaction that is reasonably realistic (e.g. 
it would apply to binaries containing neutron stars with 
baryon rest mass $1.5M_\odot$ and isolated spherical neutron stars with
a maximum baryon rest mass of $1.8M_\odot$); 
we consider the less compact model for purposes of comparison. 

We also consider mass ratios $M_{\rm irr}^{\rm BH}/M_{\rm ADM,0}^{\rm
NS}=1$, 2, 3, 5, and 10.  Note again that we fix the irreducible mass
of the black hole and the baryon rest mass of the neutron star for the
construction of constant-mass sequences.  For the definition of the
mass ratio, however, we use the ADM mass of a spherical isolated
neutron star $M_{\rm ADM,0}^{\rm NS}$, since this turns out to be more
convenient for comparisons with third-order post-Newtonian results.

We summarize our results in Tables \ref{table:seqM015} and
\ref{table:seqM01} in Appendix \ref{appendix:table}.

\begin{figure}[ht]
\vspace{0.2cm}
\begin{center}
  \includegraphics[width=8cm]{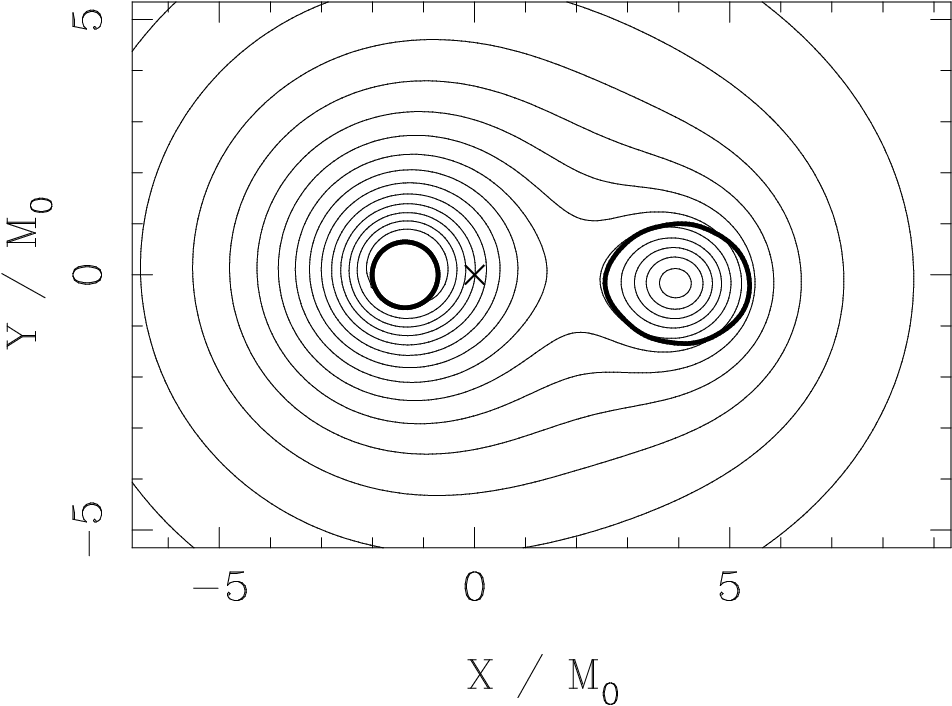}
\end{center}
\caption{Contours of the lapse function $\alpha$ in the equatorial
  plane for our innermost configuration of a binary of mass ratio
  $M_{\rm irr}^{\rm BH}/M_{\rm ADM,0}^{\rm NS}=3$ and neutron star mass
  $\bar{M}_{\rm B}^{\rm NS}=0.15$.  The minimum value of the lapse,
  $\alpha \simeq 0.453$, is located on the excised surface and the
  maximum value, $\alpha \simeq 0.909$, at the right edge of the figure.
  The value of the lapse at the point of maximum baryon rest-mass density is
  about 0.558.  The contour curves are spaced by $\delta \alpha \simeq
  0.03$.  The cross ``$\times$'' indicates the position of the rotation
  axis.}
\label{fig:lapse}
\end{figure}

In Fig.~\ref{fig:lapse} we present contours of the lapse function
$\alpha$ for a binary of mass ratio $M_{\rm irr}^{\rm BH}/M_{\rm
ADM,0}^{\rm NS}=3$ and neutron star mass $\bar{M}_{\rm B}^{\rm
NS}=0.15$ at the smallest binary separation for which our code
converged. The thick solid circle on the left-hand side marks the
position of the excised surface (the apparent horizon),
while that on the right-hand side marks the position of the neutron
star surface.
The value of the lapse function $\alpha$ on the excised surface is not
uniform because we use the Neumann boundary condition
(\ref{eq:lapse_bound}) rather than a Dirichlet boundary condition.
The approximate value of the lapse on the excision surface is
$\alpha|_{\cal S} \sim 0.45-0.48$.

\begin{figure}[ht]
\vspace{0.2cm}
\begin{center}
  \includegraphics[width=8cm]{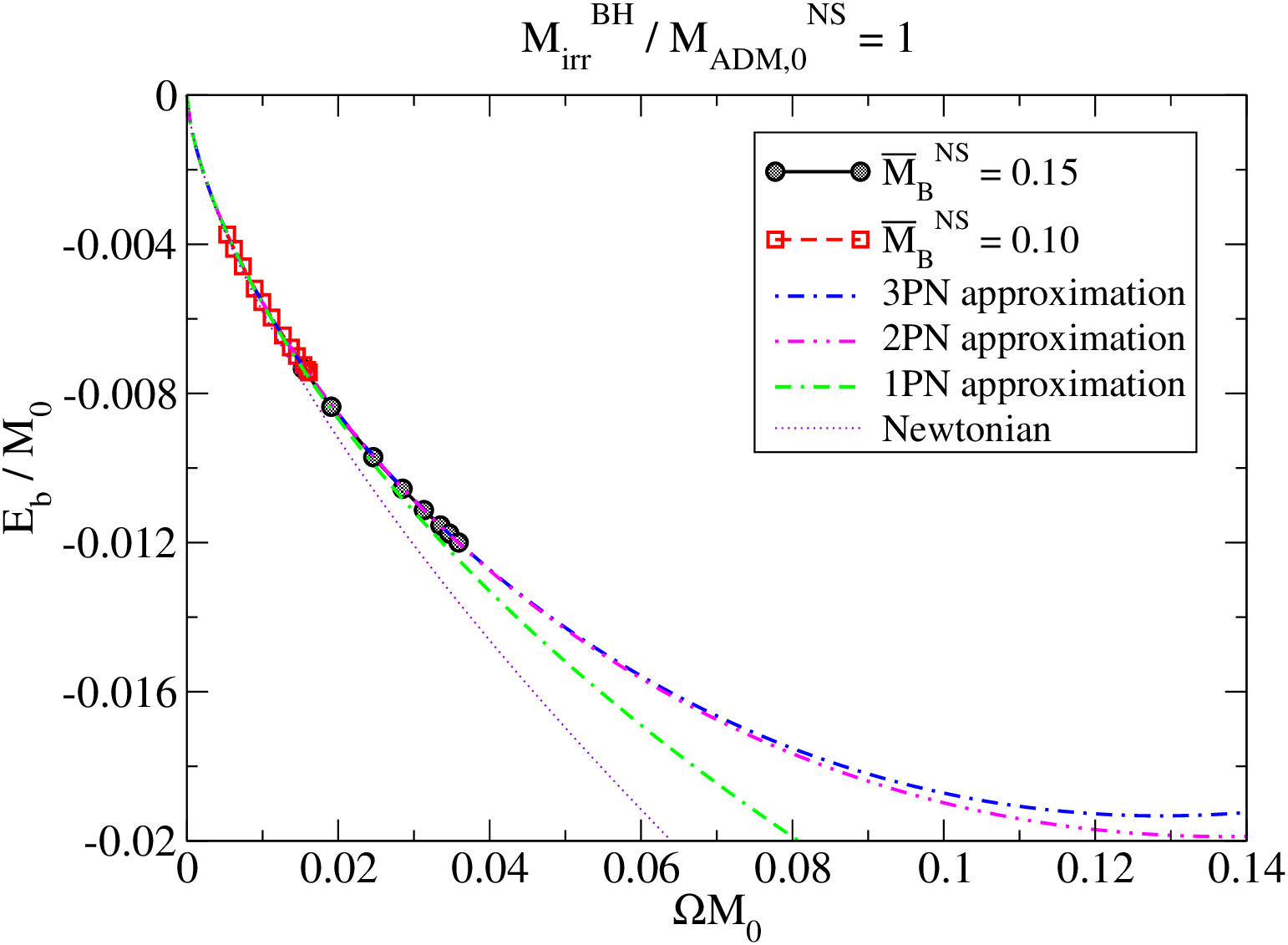}
\end{center}
\caption{Fractional binding energy $E_{\rm b}/M_0$ as a function of
  $\Omega M_0$ for binaries of mass ratio $M_{\rm irr}^{\rm
  BH}/M_{\rm ADM,0}^{\rm NS}=1$.  The solid line with filled circles
  shows results for neutron stars of mass $\bar{M}_{\rm B}^{\rm
  NS}=0.15$, and the dashed line with squares for neutron stars of
  mass of 0.10.  The dash-dotted line denotes the results of the third
  post-Newtonian approximation \cite{Blanc02}. These sequences end due
  to cusp formation -- and hence the 
  onset of tidal disruption -- before the binary reaches the ISCO at
  the turning point of the binding energy.}
\label{fig:bindene1}
\end{figure}

\begin{figure}[ht]
\vspace{0.2cm}
\begin{center}
  \includegraphics[width=8cm]{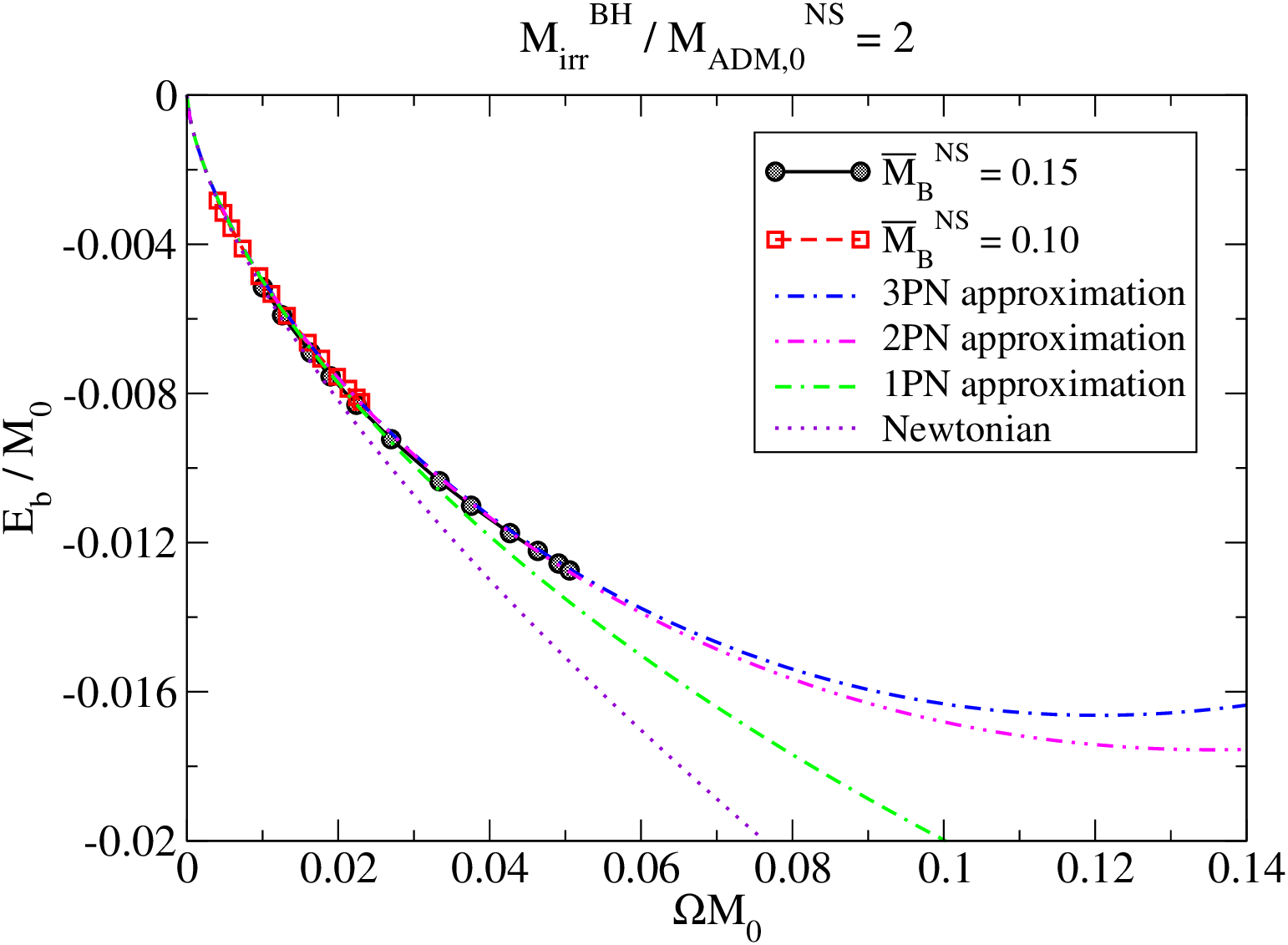}
\end{center}
\caption{Same as Fig. \ref{fig:bindene1} but for sequences of mass
  ratio $M_{\rm irr}^{\rm BH}/M_{\rm ADM,0}^{\rm NS}=2$.}
\label{fig:bindene2}
\end{figure}

\begin{figure}[ht]
\vspace{0.2cm}
\begin{center}
  \includegraphics[width=8cm]{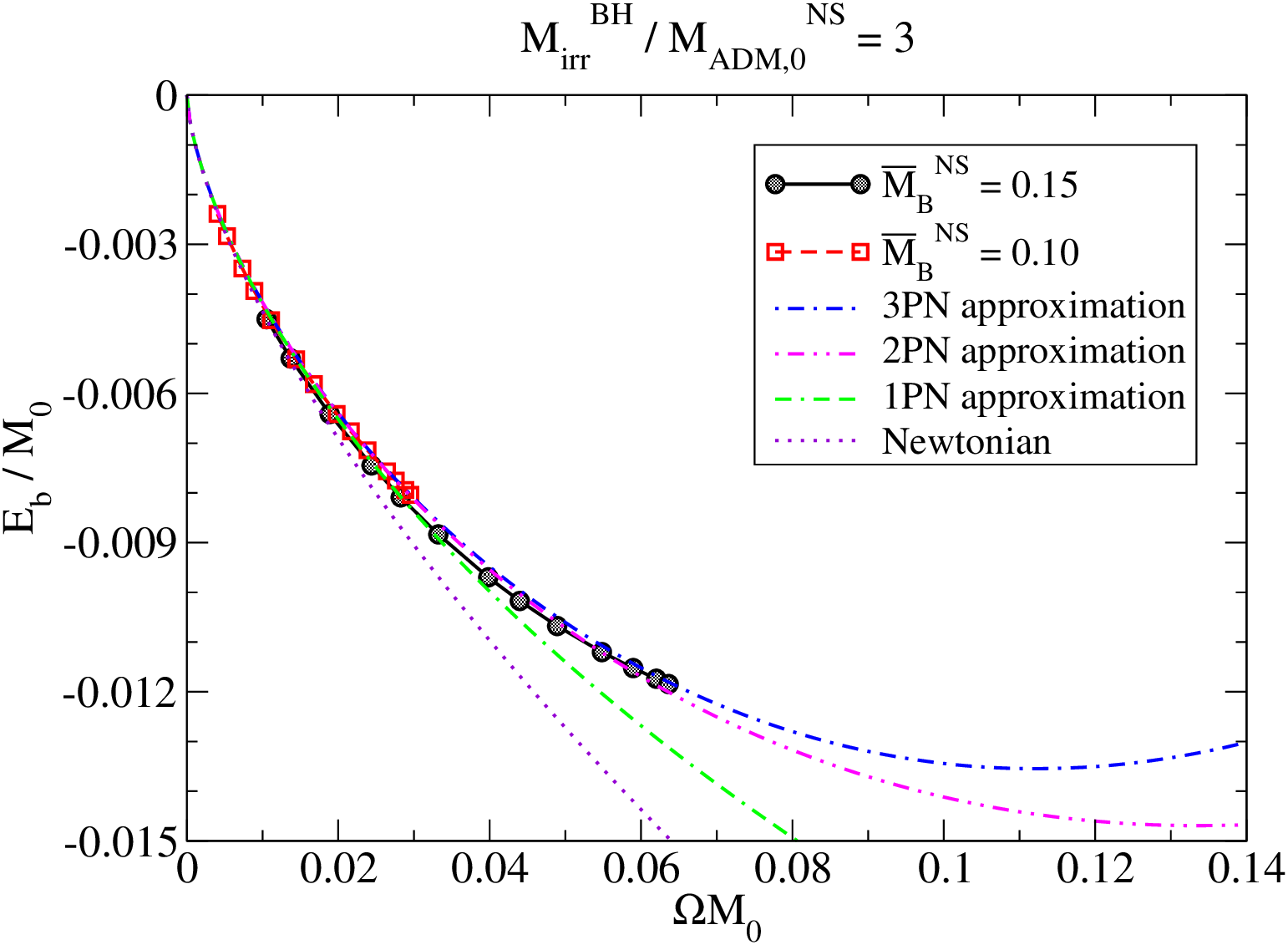}
\end{center}
\caption{Same as Fig. \ref{fig:bindene1} but for sequences of mass
  ratio $M_{\rm irr}^{\rm BH}/M_{\rm ADM,0}^{\rm NS}=3$.}
\label{fig:bindene3}
\end{figure}

\begin{figure}[ht]
\vspace{0.2cm}
\begin{center}
  \includegraphics[width=8cm]{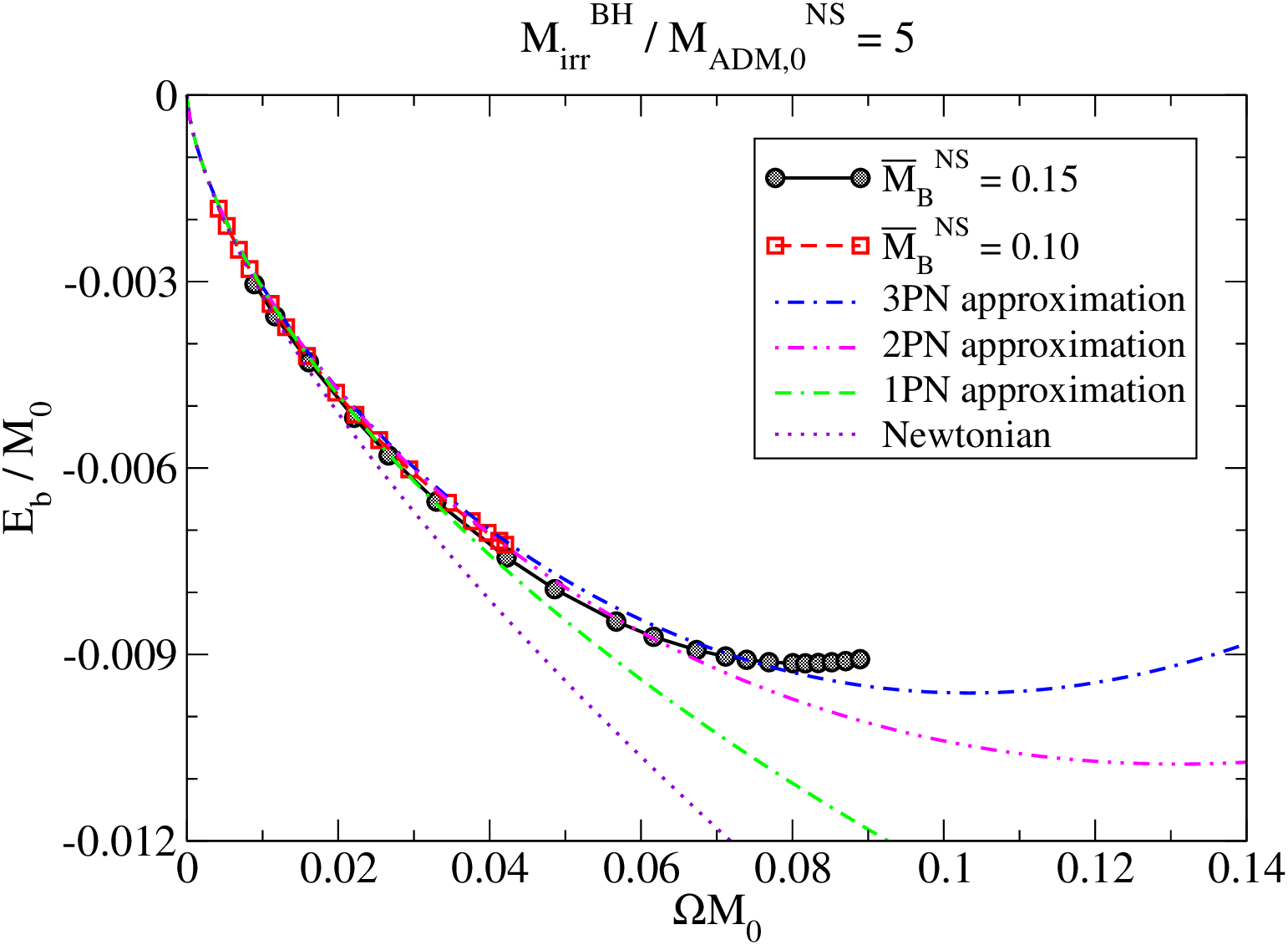}
\end{center}
\caption{Same as Fig. \ref{fig:bindene1} but for sequences of mass
  ratio $M_{\rm irr}^{\rm BH}/M_{\rm ADM,0}^{\rm NS}=5$.  For the more
  compact neutron star the binary now encounters an ISCO before the
  neutron star is tidally disrupted.}
\label{fig:bindene5}
\end{figure}

\begin{figure}[ht]
\vspace{0.2cm}
\begin{center}
  \includegraphics[width=8cm]{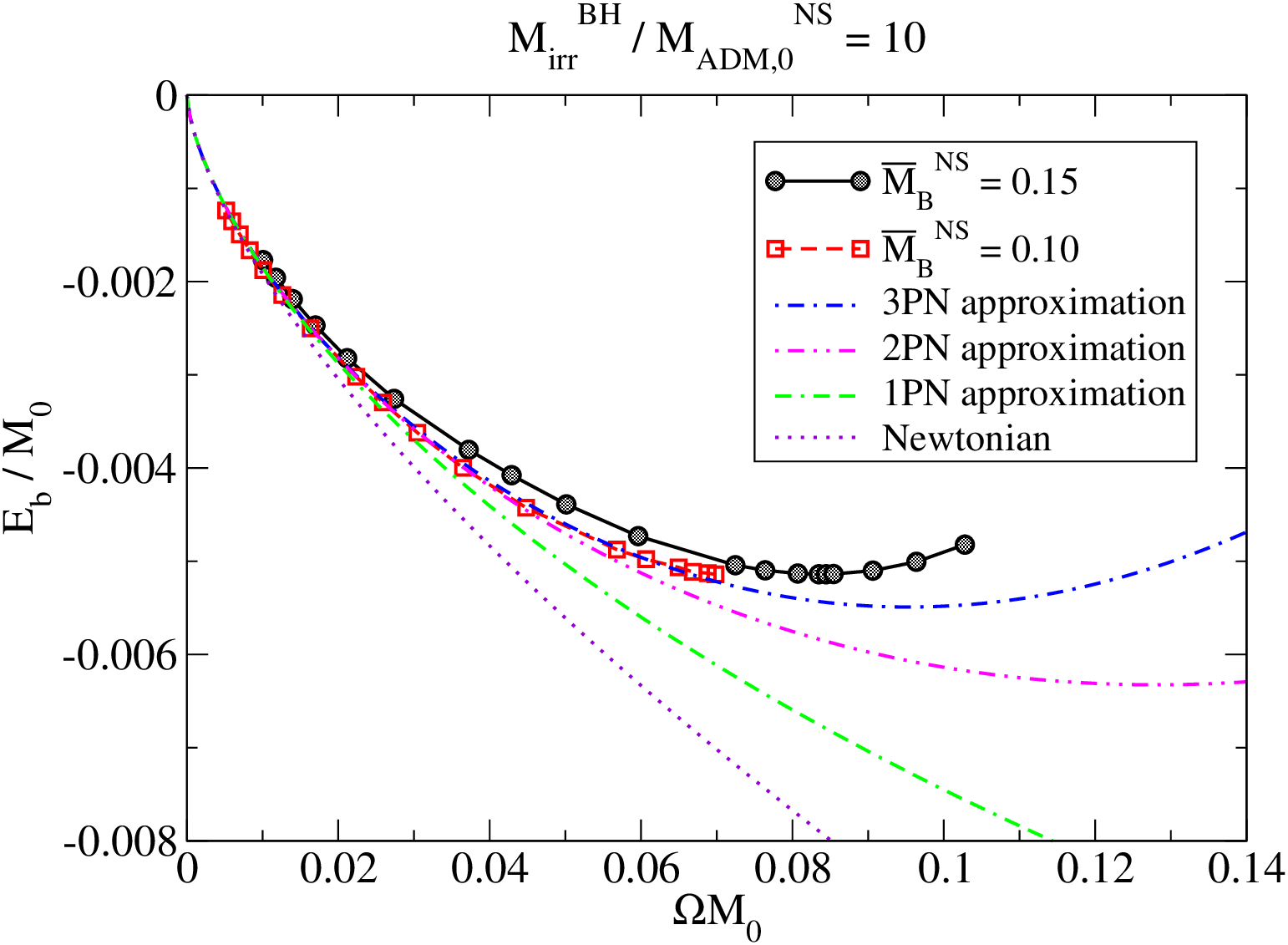}
\end{center}
\caption{Same as Fig. \ref{fig:bindene1} but for sequences of mass ratio
  $M_{\rm irr}^{\rm BH}/M_{\rm ADM,0}^{\rm NS}=10$.}
\label{fig:bindene10}
\end{figure}

In Figs.~\ref{fig:bindene1} -- \ref{fig:bindene10} we plot the binding
energy $E_{\rm b}/M_0$ versus the orbital angular velocity $\Omega
M_0$ for sequences with mass ratios 1, 2, 3, 5 and 10.  We find
excellent agreement
with third-order post-Newtonian approximations \cite{Blanc02},
especially for large binary separations.  For more compact neutron
stars with $\bar{M}_{\rm B}^{\rm NS}=0.15$, we also find a turning
point in the sequences with mass ratios $M_{\rm irr}^{\rm BH}/M_{\rm
ADM,0}^{\rm NS}=5$ and 10.  Such turning points mark the ISCO,
inside of which no stable circular orbits can exist.  For irrotational
binaries this instability is dynamical (see, e.g.~\cite{LaiRS93}).
Binaries with larger neutron star compaction and larger mass ratios
therefore encounter an ISCO before being tidally disrupted.
For all other binaries, the sequences end due to cusp
formation -- and hence at the onset of tidal disruption -- before the
binary encounters an ISCO.

Qualitatively, this behavior can be understood very easily from a
simple Newtonian analysis.  The binary separation at the onset of
tidal disruption, $d_{\rm tid}$, can be estimated by equating the
gravitational and tidal forces on a test particle on the surface of
the neutron star
\begin{equation} \label{isco_scaling}
  \frac{d_{\rm tid}}{M_{\rm BH}} \sim
  \left( \frac{M_{\rm NS}}{M_{\rm BH}} \right)^{2/3}
  \frac{R_{\rm NS}}{M_{\rm NS}}.
\end{equation}
If this is smaller than the binary separation at the ISCO, at $d_{\rm
ISCO}/M_{\rm BH} = 6$ for a test particle in Schwarzschild
coordinates, the binary encounters the ISCO before being tidally
disrupted.  This happens for large mass ratios $M_{\rm BH}/M_{\rm NS}$
and large compactions $M_{\rm NS}/R_{\rm NS}$ -- in accordance with
our findings.

In the test-particle limit the location of the turning point should
coincide with that of test particle orbiting a Schwarzschild black
hole of mass $M_0$, $\Omega M_0 = 6^{-3/2} \simeq 0.068$.
We find similar but slightly larger
values, $\Omega M_0 \sim 0.082$ and $\sim 0.084$ for $M_{\rm irr}^{\rm
BH}/M_{\rm ADM,0}^{\rm NS}=5$ and 10.  Locating the minimum of a
numerically generated curve always introduces some additional error,
and it is possible that this deviation is entirely a numerical
artifact.  However, it is also possible that the deviation is a
consequence of our ``leading-order approximation'' of the black hole
spin.  This approximation leads to a small but non-zero black hole
spin, so that effectively we locate the ISCO around a Kerr instead of
a Schwarzschild black hole.  Further evidence for this hypothesis
is the fact that we do not find turning points of the angular momentum
that coincide with those of the binding energy.  In \cite{CookP04},
the authors similarly found that the binding energy and angular
momentum for ``leading-order'' irrotational black hole binaries did
not have simultaneous turning points.  In \cite{CaudiCGP06}, however,
the authors show that an improved condition for nonspinning black
holes leads to a very good agreement between the binding energy and
angular momentum turning points.  We plan to implement this improved
condition in the near future.

\begin{figure}[ht]
\vspace{0.3cm}
\begin{center}
  \includegraphics[width=8cm]{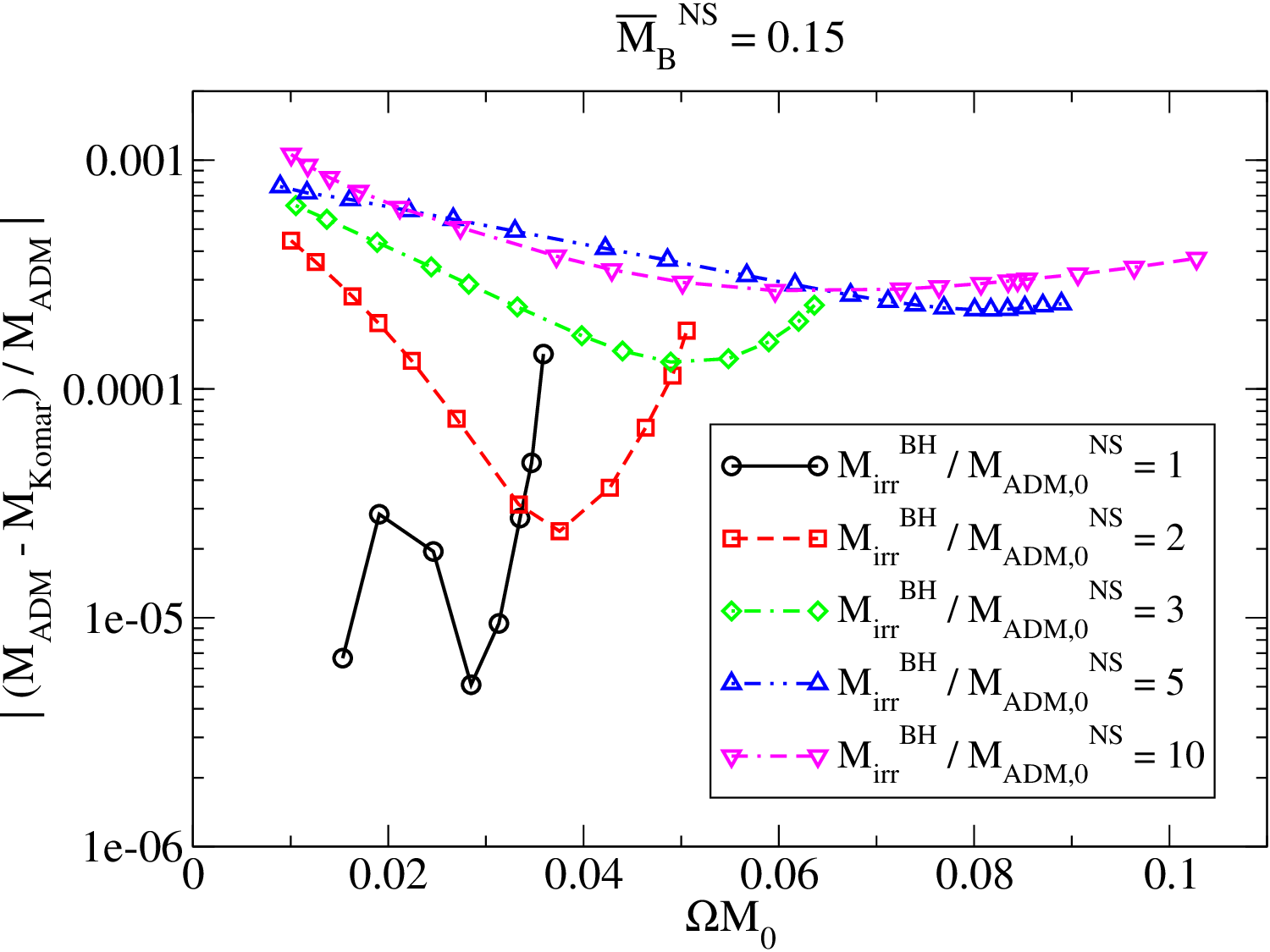} \\
\vspace{0.4cm}
  \includegraphics[width=8cm]{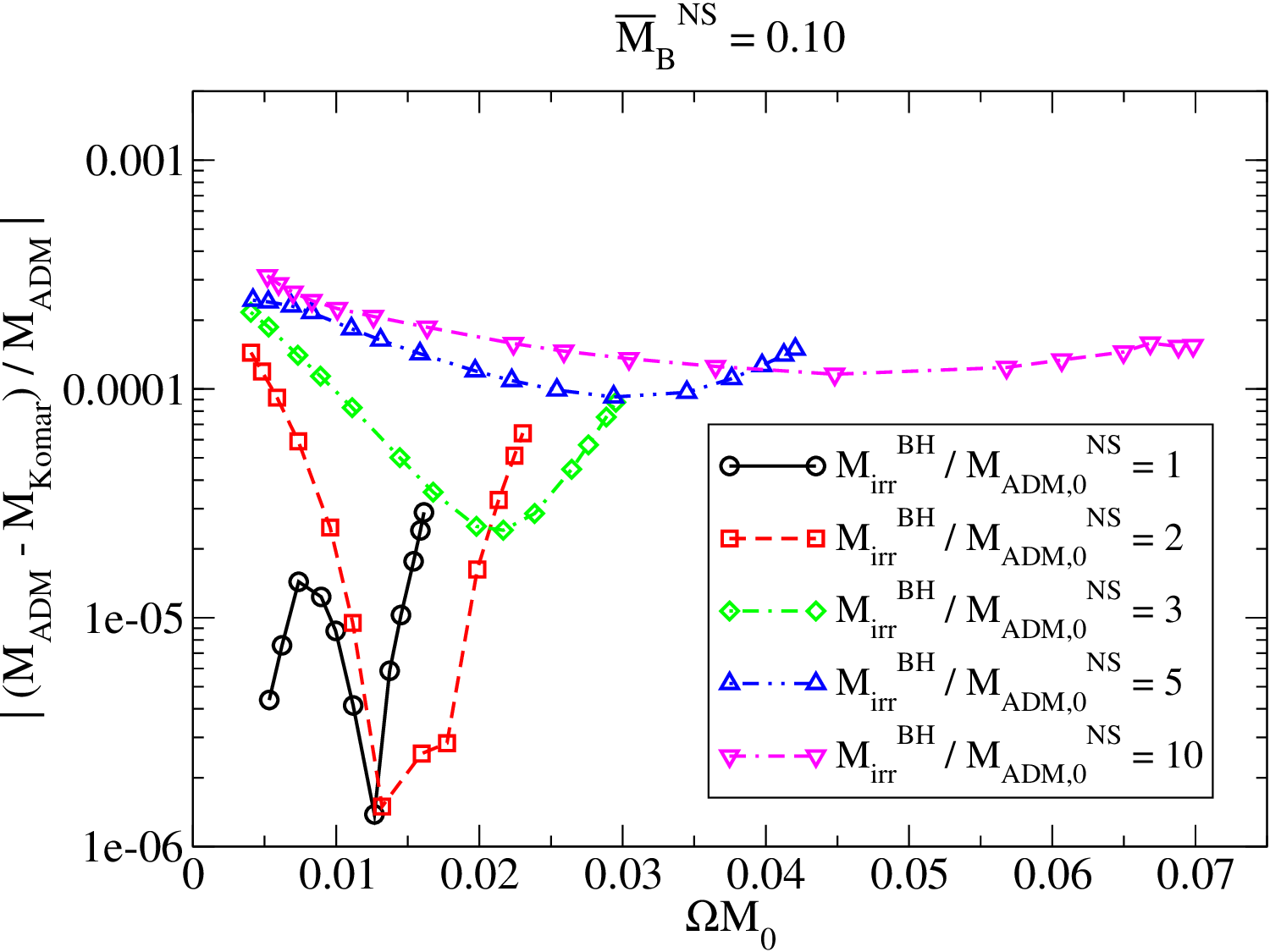}
\end{center}
\caption{Virial error $\delta M$ versus $\Omega M_0$ for our binary
  sequences.}
\label{fig:virial}
\end{figure}

As one measure of the error in our numerical calculations we 
compute the virial expression (\ref{eq:virial}).  As shown in 
Fig.~\ref{fig:virial}, all our configurations have relative virial
errors of less than $10^{-3}$, generally on the order of several
$\times 10^{-4}$.  The errors are smaller at intermediate binary
separations, and somewhat larger both at larger binary separations (where
the angular resolution becomes worse) and smaller binary separations
(where the larger tidal deformation of the neutron star causes larger
numerical error).

 From the virial error we can estimate the error in the binding 
energy as follows.
We assume that the relative error in the ADM mass 
is of the same order as the relative virial error,
\begin{equation}
  \Bigl| {M_{\rm ADM} - M_{\rm ADM,true} \over M_{\rm ADM,true}}
  \Bigr| \sim \delta M, \label{eq:errorADM}
\end{equation}
where $M_{\rm ADM,true}$ is the true value of the ADM mass.
 From the definition of the binding energy
Eq.~(\ref{eq:bindene}), we can write the relative difference of the
binding energy from its true value $E_{\rm b,true}$ as
\begin{equation}
  \Bigl| {E_{\rm b} - E_{\rm b,true} \over E_{\rm b,true}} \Bigr| \sim
  \delta M \Bigl| {M_{\rm ADM,true} \over M_{\rm ADM,true} -M_0}
  \Bigr|. \label{eq:error}
\end{equation}
Since the term $|M_{\rm ADM,true}/(M_{\rm ADM,true} -M_0)|$ on the
right-hand side is of order of $10^2$ for the binary separations that 
we calculate, we conclude that the relative error in the
binding energy is approximately $10^2 \delta M$.  For our least
accurate sequences this results in a relative error of several $\times
10^{-2}$ and for our most accurate sequences it is a few $\times
10^{-3}$.

\begin{figure}[ht]
\vspace{0.3cm}
\begin{center}
  \includegraphics[width=8cm]{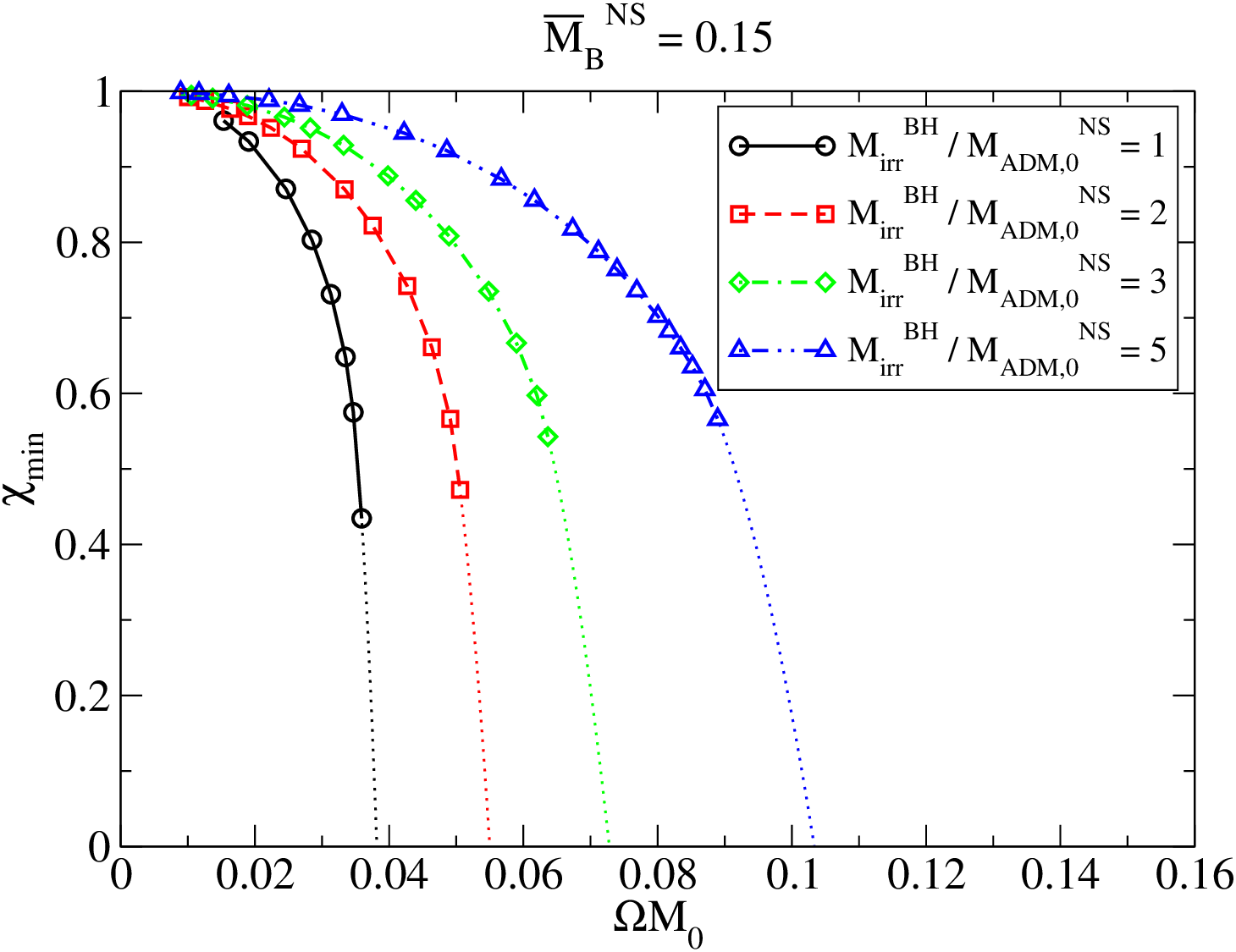} \\
\vspace{0.4cm}
  \includegraphics[width=8cm]{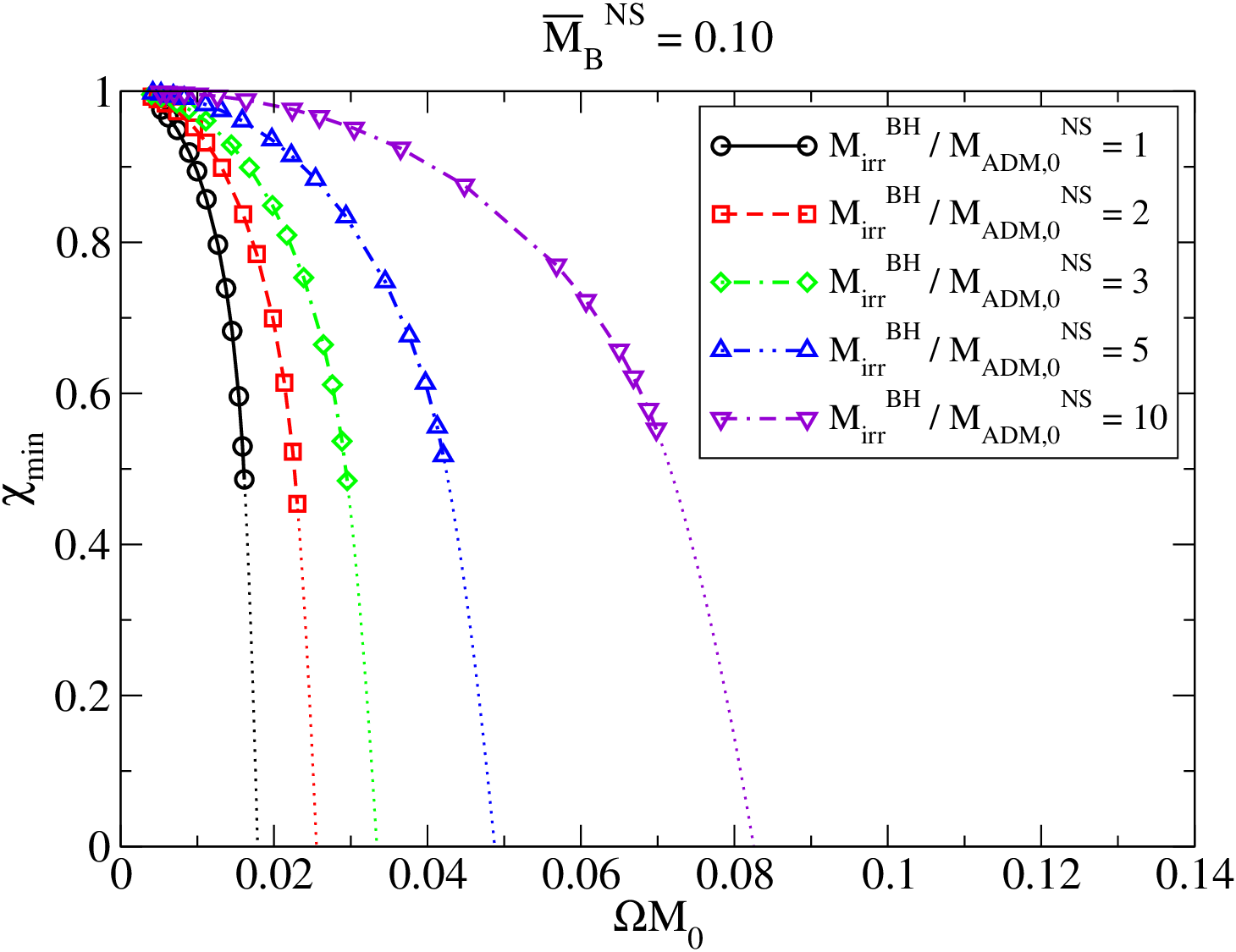}
\end{center}
\caption{Minimum of the mass-shedding indicator $\chi_{\rm min}$ 
  as a function of $\Omega M_0$.}
\label{fig:chi}
\end{figure}

\begin{table}[ht]
\caption{Estimated orbital angular velocity at the tidal disruption
  point.}
\begin{center}
\begin{tabular}{c|l|l} \hline\hline
  &\multicolumn{2}{c}{$\Omega M_0$} \\
  $M_{\rm irr}^{\rm BH}/M_{\rm ADM,0}^{\rm NS}$&
  $\bar{M}_{\rm B}^{\rm NS}=0.10$&$\bar{M}_{\rm B}^{\rm NS}=0.15$ \\
  \hline
  1 & 0.0179 & 0.0382 \\
  2 & 0.0255 & 0.0550 \\
  3 & 0.0334 & 0.0728 \\
  5 & 0.0488 & 0.103 \\
  10 & 0.0825 & ------ \\ \hline
\end{tabular}
\end{center}
\label{table:omega}
\end{table}

Using a spectral code we cannot follow our sequences all the way to
cusp formation (see our discussion in \cite{TanigBFS05}).  To monitor
the formation of a cusp we therefore define a mass-shedding indicator
\cite{GourgGTMB01}
\begin{equation}
  \chi \equiv {(\partial (\ln h)/\partial r)_{\rm eq}
    \over (\partial (\ln h)/\partial r)_{\rm pole}},
\end{equation}
the ratio of the radial derivative of the enthalpy in the equatorial
plane at the surface to that to the polar direction at the surface.
Given our choice of boundary conditions the minimum point of $\chi$ is
not on the $X$ axis (which connects the neutron star with the black
hole), but slightly away from the axis.  In Fig.~\ref{fig:chi} we
therefore graph the minimum value of $\chi$, searched on the surface
of the neutron star, as a function of $\Omega M_0$.  For spherical
stars at infinite separation we have $\chi_{\rm min}=1$, while
$\chi_{\rm min}=0$ indicates the formation of a cusp and hence tidal
breakup.  Our spectral code stops converging before reaching
$\chi_{\rm min}=0$, but the sharp drop in $\chi_{\rm min}$ is an
indication of the formation of a cups.  Extrapolating to $\chi_{\rm
min}=0$ from the last three points (indicated by the thin dotted lines
in
Fig.~\ref{fig:chi}) we estimate the orbital angular velocity at the
onset of tidal disruption (see Table \ref{table:omega}).  We did not
include the sequence with mass ratio
$M_{\rm irr}^{\rm BH}/M_{\rm ADM,0}^{\rm NS}=10$ and neutron star
mass $\bar{M}_{\rm B}^{\rm NS}=0.15$
in this analysis, since the binary encounters an ISCO long before the
neutron star is tidally disrupted (see also Table \ref{table:seqM015}).
For the sequence with $M_{\rm irr}^{\rm BH}/M_{\rm ADM,0}^{\rm NS}=10$
and $\bar{M}_{\rm B}^{\rm NS}=0.10$ the extrapolation of $\chi_{\rm min}$ 
suggests that tidal disruption occurs at $\Omega M_0 \simeq 0.0825$
(the last entry in Table \ref{table:omega}), a value very similar to 
the values of the ISCO in those two sequences for which we found
a turning point.  This suggests that the neutron star in this binary
may be tidally disrupted just as the orbit becomes unstable.

\begin{figure}[ht]
\vspace{0.3cm}
\begin{center}
  \includegraphics[width=8cm]{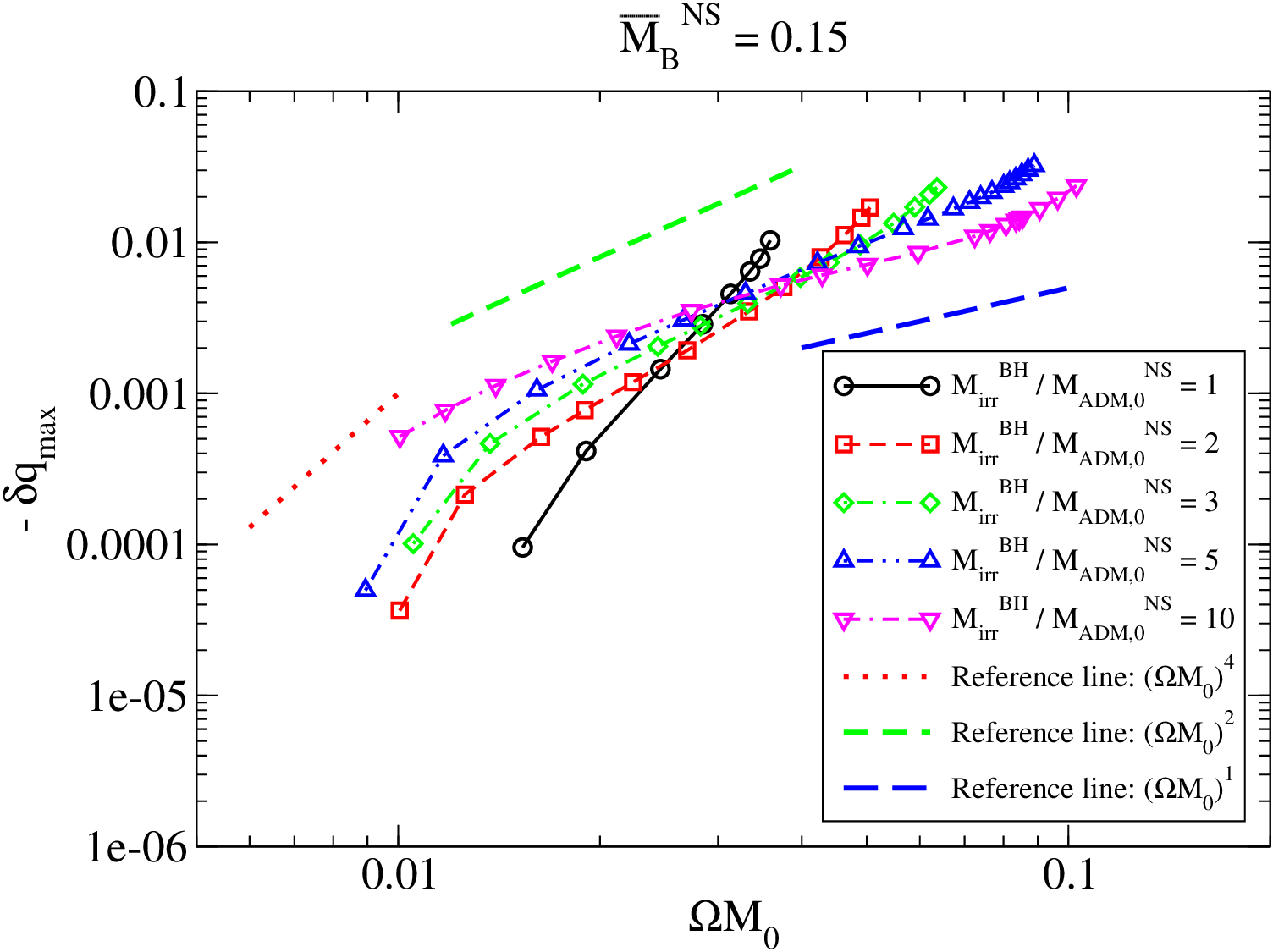} \\
\vspace{0.4cm}
  \includegraphics[width=8cm]{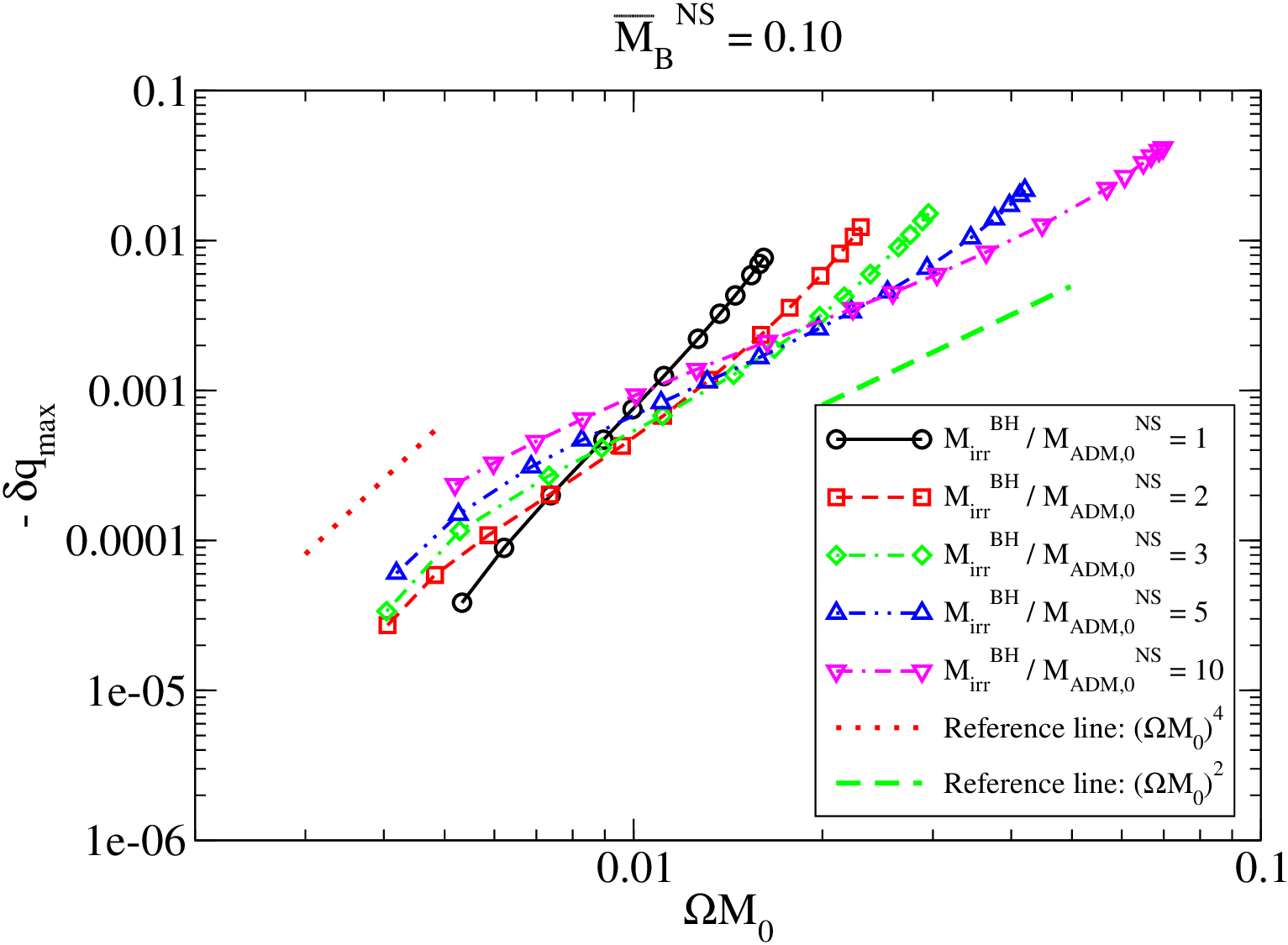}
\end{center}
\caption{Decrease in the maximum density parameter
  $\delta q_{\rm max}$ as a function of $\Omega M_0$.}
\label{fig:dqmax}
\end{figure}

In \cite{TanigBFS06}, where we computed similar BHNS binary sequences
but with a Kerr-Schild background, we also considered a binary of mass
ratio $M_{\rm irr}^{\rm BH}/M_{\rm B}^{\rm NS}=5$ and neutron star
mass $\bar{M}_{\rm B}^{\rm NS}=0.10$.  There we estimated that tidal
disruption would set in at $\Omega M_0 \simeq 0.046$.  We compute
a sequence with the same definition of the mass ratio
$M_{\rm irr}^{\rm BH}/M_{\rm B}^{\rm NS}=5$ for comparison, and
find tidal disruption at approximately $\Omega M_0 \sim 0.053$,
a value within 15\% of our earlier value. 

As the orbital separation decreases, the maximum density of the 
neutron star decreases.  
We define the dimensionless density parameter 
\begin{equation}
  q \equiv \frac{P}{\rho_0}
\end{equation}
and monitor the decrease in the maximum density
\begin{equation}
  \delta q_{\rm max} \equiv {q_{\rm max} - q_{\rm max,0} \over
    q_{\rm max,0}},
\end{equation}
where $q_{\rm max,0}$ denotes that of an isolated neutron star with the
same baryon rest mass.   For 
$\bar{M}_{\rm B}^{\rm NS}=0.15$ (0.10) we have 
$q_{\rm max,0}=0.12665$ (0.058827).
Note that $\delta q_{\rm max}$ is
negative for all configurations considered in this study. 
In Fig.~\ref{fig:dqmax} we therefore graph 
$-\delta q_{\rm max}$ as a function of the orbital angular velocity
$\Omega M_0$ on a logarithmic scale.  

Within Newtonian gravity, the decrease in the central density
scales, to leading order, with
\begin{equation}
  \delta q \propto d^{-\sigma} \propto \Omega^{2\sigma/3},
\end{equation}
where $d$ is the orbital separation,
and for irrotational binaries the index $\sigma$ takes the value
$\sigma = 6$ (see, e.g.~\cite{TanigN00} for an analytic derivation and
\cite{GourgGTMB01} for comparison with numerical results).  Here the
leading order term is caused by a quadrupole deformation in response
to the companion's tidal field.  We see from Fig.~\ref{fig:dqmax}
that the power-law index $\sigma$ is indeed close to 6 for large binary 
separations.  For intermediate  binary separations, however, the index
is generally smaller, $\sigma\sim 2-3$.  For the equal-mass binary
$\sigma$ remains close to 6 for the entire sequence, but for
increasing mass ratios $\sigma$ decreases at intermediate binary
separations.  We plan to investigate this behavior with the help
of semi-analytic models in a future publication.
Finally, for small binary separations, just before
tidal disruption sets in, $\sigma$ increases again.  This has also
been seen for BNS systems \cite{TanigG02,TanigG03} and may be
caused by higher order multipole deformations. 

\begin{figure}[ht]
\vspace{0.3cm}
\begin{center}
  \includegraphics[width=8cm]{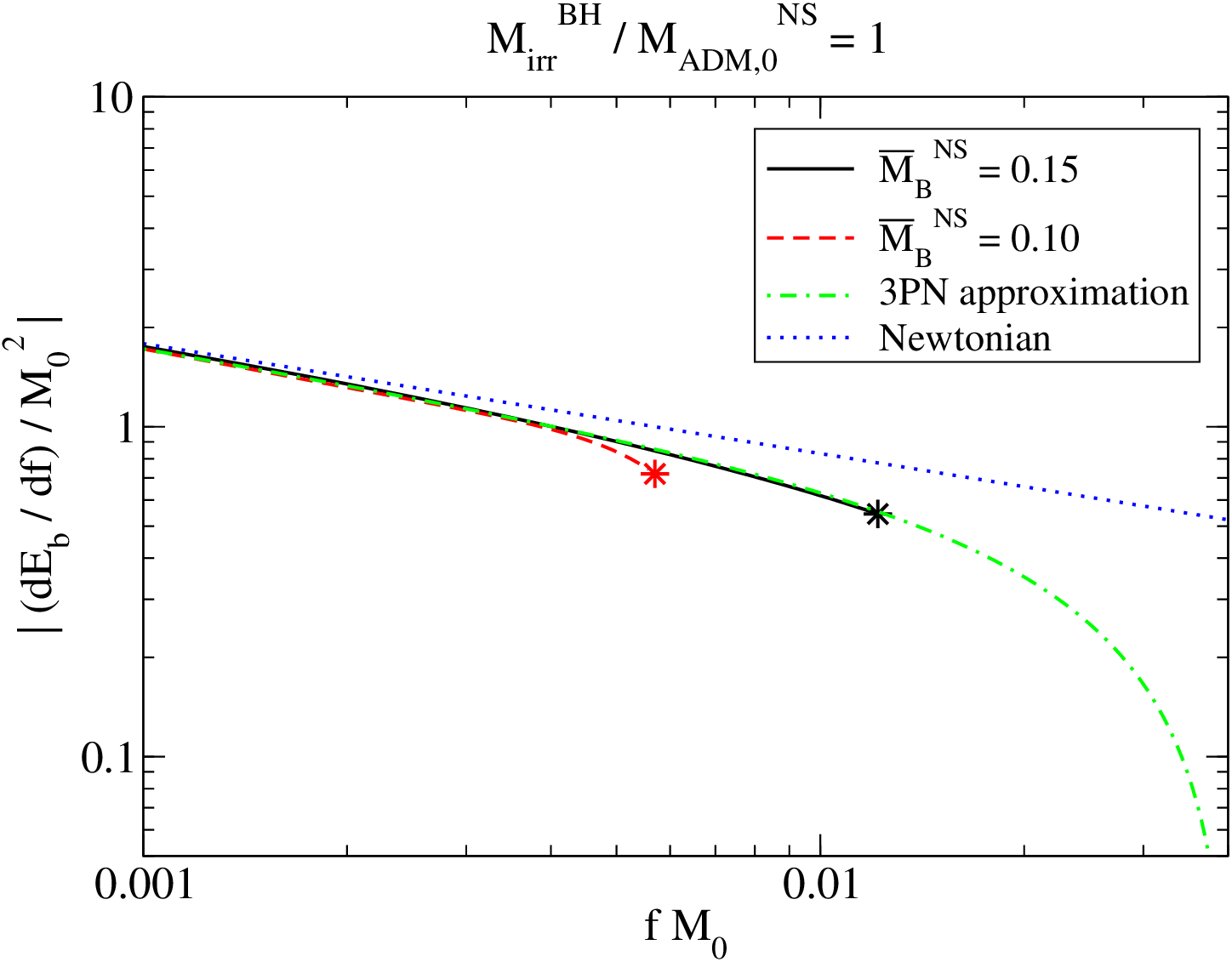}
\end{center}
\caption{The energy spectrum
  $dE_{\rm b}/df$ for the sequences with
  $M_{\rm irr}^{\rm BH}/M_{\rm ADM,0}^{\rm NS}=1$
  as a function of the dimensionless
  gravitational wave frequency 
  $f M_0$.  The solid, dashed, dash-dotted, and dotted curves
  show the fits to the $\bar{M}_{\rm B}^{\rm NS}=0.15$ and
  $\bar{M}_{\rm B}^{\rm NS}=0.1$ sequences and the 3PN and
  Newtonian expressions, respectively.  Asterisks denote the
  approximate point where a cusp forms and tidal disruption begins, 
  terminating the sequence, at
  frequencies given by Table \protect\ref{table:omega}.}
\label{fig:dedf1}
\end{figure}

\begin{figure}[ht]
\vspace{0.3cm}
\begin{center}
  \includegraphics[width=8cm]{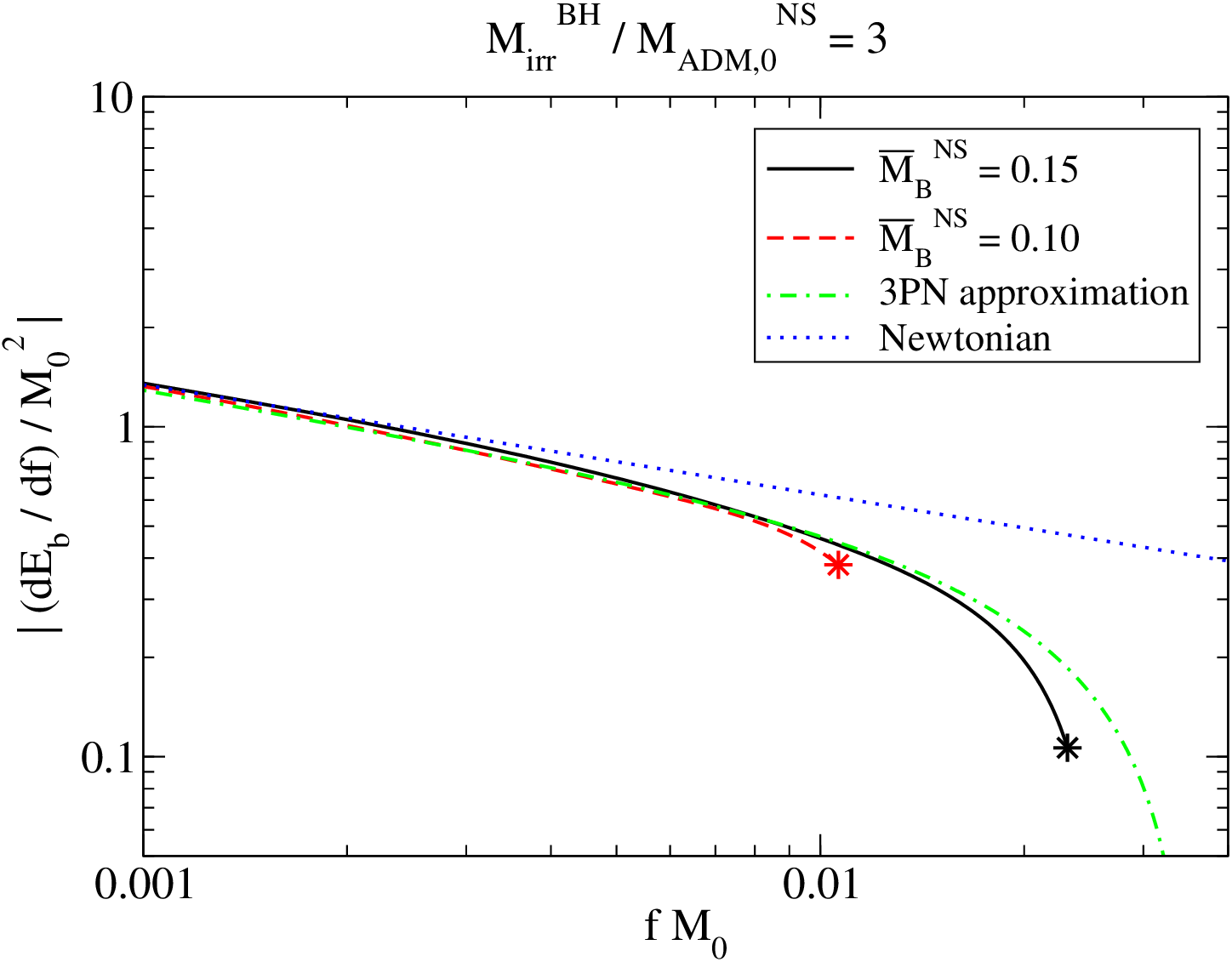}
\end{center}
\caption{Same as Fig.~\protect\ref{fig:dedf1}, but for the sequences
  with $M_{\rm irr}^{\rm BH}/M_{\rm ADM,0}^{\rm NS}=3$.}
\label{fig:dedf3}
\end{figure}

\begin{figure}[ht]
\vspace{0.3cm}
\begin{center}
  \includegraphics[width=8cm]{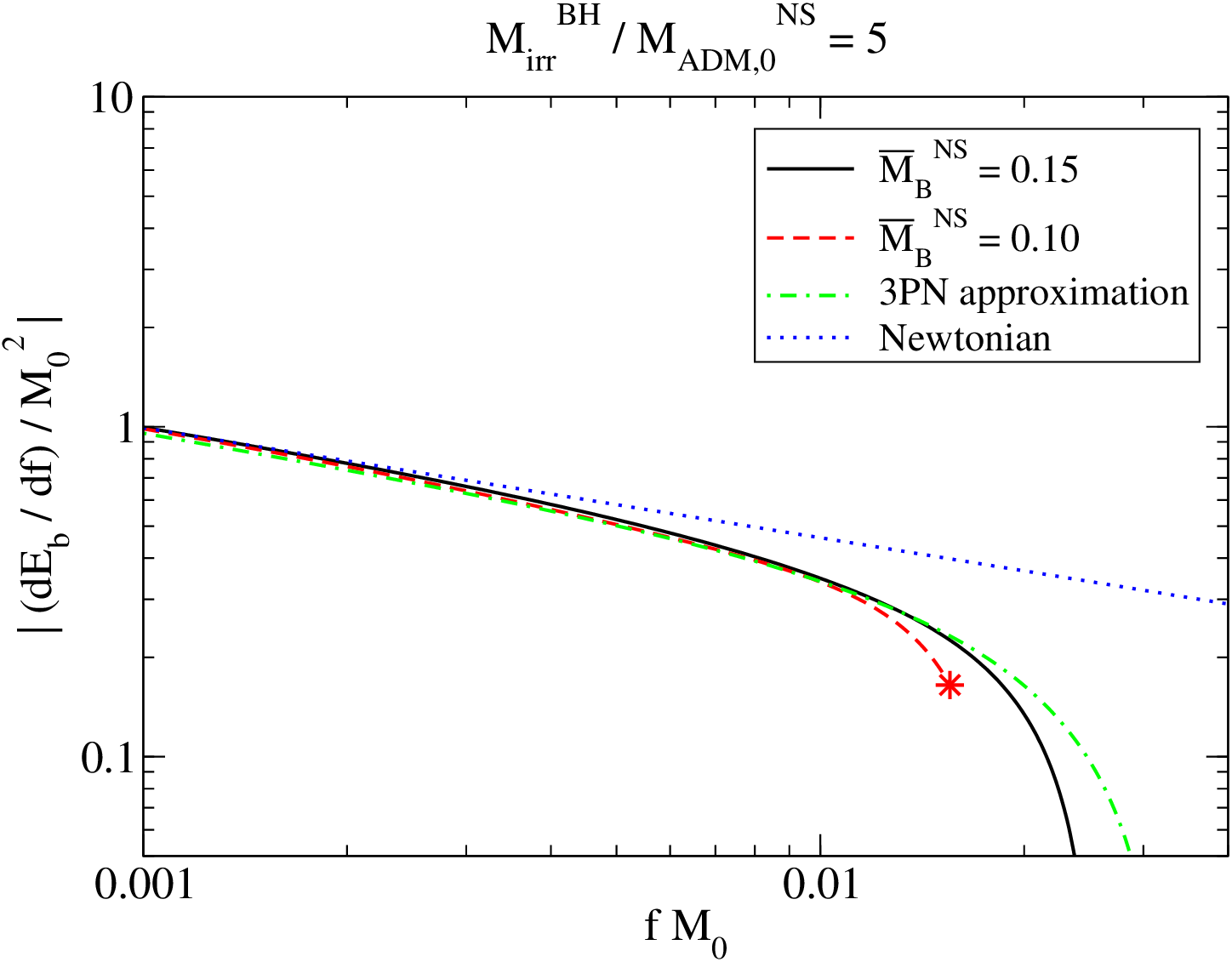}
\end{center}
\caption{Same as Fig.~\protect\ref{fig:dedf1}, but for the sequences
  with $M_{\rm irr}^{\rm BH}/M_{\rm ADM,0}^{\rm NS}=5$.  Note that the
  sequence with $\bar{M}_{\rm B}^{\rm NS}=0.15$ reaches a minimum
  binding energy, where $dE_{\rm b}/df=0$, prior to tidal disruption.}
\label{fig:dedf5}
\end{figure}

Finally, we turn our attention to the energy spectrum of
gravitational waves emitted from a BHNS
binary, which is calculated from the Fourier transform of the
gravitational wave strains averaged over all angles
(see, e.g.~\cite{Zhuge94}),  
\begin{equation}
  \frac{dE}{df}=\frac{\pi}{2}(4\pi
  r^2)f^2\left<|\tilde{h}_+(f)|^2+
  |\tilde{h}_\times(f)|^2\right>, 
\end{equation}
where the brackets denote angle averaging and
\begin{equation}
  \tilde{h}(f)=\int_{-\infty}^\infty h(t)e^{2\pi ift}dt,
\end{equation}
where $f$ is the gravitational wave frequency, rather than the orbital
frequency. From the results of our quasiequilibrium sequences, we can
compute approximate gravitational wave energy spectra using the
methods described in \cite{FaberGRT02}.  To do so, we assume that
gravitational waves from the binary are emitted coherently with
$f=\Omega/\pi$ at any given moment in
time, where $\Omega$ is the standard orbital angular velocity, and further 
assume that the system is adiabatic and infall velocities negligible, 
so that the gravitational wave energy losses drive the system
gradually through a series of quasiequilibrium configurations. 
Under these assumptions, we recover a simple expression for the
gravitational wave energy spectrum,
\begin{equation}
  \frac{dE}{df} \approx
  -\pi \left( \frac{dE_{\rm b}}{d\Omega}\right)_{\rm QE},
\end{equation}
i.e., we can calculate the energy spectrum by numerically
differentiating the binding energy along the quasiequilibrium sequence
with respect to the angular frequency. This method has been shown
to reproduce closely the actual energy spectrum calculated from a
fully dynamical binary neutron star merger calculation in the
frequency regime of interest \cite{FaberGR04}.  While we can
numerically extrapolate up to the frequency at which we find a minimum
in the binding energy, 
representing the ISCO, our results do not apply inside the separation
at which a cusp forms and tidal disruption begins (see Table
\ref{table:omega}).  Dynamically, this
would lead to the beginning of mass transfer and a rapid evolution
away from quasiequilibrium, which in turn would produce a
qualitatively different gravitational wave signal in both the time and
frequency domains.

Instead of taking derivatives of our numerical data for the binding 
energy directly (see Tables \ref{table:seqM015} and \ref{table:seqM01}), 
we first fit these data to a fourth-order polynomial of the
form 
\begin{equation}
  \frac{E_{\rm b}}{M_0}=a_{\rm N} x
  \left(1+a_1 x+a_2x^2+a_3x^3+a_4x^4\right)
  \label{eq:fit}
\end{equation}
where $x\equiv (\Omega M_0)^{2/3}=(\pi
f M_0)^{2/3}$.  Given this parametrization of the polynomial,
the coefficients $a_i$ can be interpreted as post-Newtonian terms of
order $i$.  The leading order, Newtonian term, is given by
\begin{equation}
  a_{\rm N}=-\frac{\nu}{2}\equiv
  -\frac{M_{\rm irr}^{\rm BH} M_{\rm ADM,0}^{\rm NS}}{2M_0^2}.
\end{equation}
Here $\nu$ is the symmetric mass ratio, i.e., the ratio of the reduced
mass of the system to the total mass.  
We tabulate the coefficients $a_i$, together with the exact 3PN
coefficients for point-mass systems as listed in \cite{Blanc02}, for
the different binary sequences in Table~\ref{table:dedf}.  We also
include the ``cutoff frequencies'' $f_n$, denoting
the frequency at which the spectral energy is $n$\% of its Newtonian
counterpart,
\begin{eqnarray}
  \frac{1}{M_0^2}\frac{dE_{\rm b}}{df}(f_n) &=&
  \frac{n}{100}\times \left(\frac{1}{M_0^2}
  \frac{dE_{\rm b}}{df}\right)_{\rm N} \nonumber \\
  &=&\frac{n}{100} \frac{2\pi a_{\rm N}}{3}
  \left(\pi f_n M_0\right)^{-1/3},
\label{eq:omegacut}
\end{eqnarray}
for $n= 75, 50$ and 25. 

In Figs. \ref{fig:dedf1} -- \ref{fig:dedf5}, we show the results of
the energy spectrum of gravitational waves.
As is expected, binaries of comparable mass terminate with the
formation of a cusp, rather than at an ISCO denoting a minimum in the
binding energy of the binary.  For the smallest black hole mass we consider, 
$M_{\rm irr}^{\rm BH}/M_{\rm ADM,0}^{\rm NS}=1$, shown in
Fig.~\ref{fig:dedf1}, tidal disruption occurs
before we see significant deviations from either 
the lowest-order Newtonian point-mass results or the exact 3PN
point-mass expression. 
As the black hole mass and tidal disruption frequency both increase,
so do the deviations in the gravitational wave energy spectrum away
from the Newtonian point-mass result.  This is especially true for the more
compact neutron star models with $\bar{M}_{\rm B}^{\rm NS}=0.15$.  For a mass
ratio $M_{\rm irr}^{\rm BH}/M_{\rm ADM,0}^{\rm NS}=5$, shown in
Fig.~\ref{fig:dedf5}, the less compact neutron star model with $\bar{M}_{\rm
  B}^{\rm NS}=0.1$ (dashed curve) shows significant deviations from the
3PN result at the tidal disruption point, whereas the more compact
model with $\bar{M}_{\rm B}^{\rm NS}=0.15$ (solid curve) does reach
the ISCO before the onset of tidal disruption.  This is also the
case for the most extreme mass ratio we consider, $M_{\rm
  irr}^{\rm BH}/M_{\rm ADM,0}^{\rm NS}=10$, for which any physically
reasonably compact neutron star model reaches an ISCO prior to the tidal
disruption point, as had been confirmed by our previous dynamical
calculations \cite{FaberBST06}.

We note that the gravitational wave spectrum should produce an
important constraint on the neutron star compaction, regardless of how the
sequence terminates.  As the compaction
decreases, tidal effects grow in importance, and tend to increases the
binding energy of the system (lowering the magnitude of the negative
binding energy), especially at smaller separations, implying that the
gravitational wave spectrum will be cutoff at lower frequencies.

\begin{table}[ht]
\caption{Exact 3PN data from \protect\cite{Blanc02}, along with
  polynomial fit results from the sequences with neutron star masses
  $\bar{M}_{\rm B}^{\rm NS}=0.15$ and 0.10.  Coefficients
  are defined by Eq.~(\protect\ref{eq:fit}).  Characteristic
  gravitational wave cutoff frequencies $f_{75} M_0$,
  $f_{50}M_0$, and $f_{25} M_0$
  are defined by Eq.~(\protect\ref{eq:omegacut}).  Cutoff frequencies
  shown in boldface are interpolated from our data, while those in
  regular text are
  extrapolated.  Values shown in italics occur at separations within our best
  estimate of the tidal disruption point, given by the values in Table
  \protect\ref{table:omega}, and likely overestimate the true cutoff
  frequency.}
\begin{tabular}{l|cccc|ccc}
\hline\hline
  & $a_1$ & $a_2$ & $a_3$ & $a_4$ &
  $f_{75} M_0$ & $f_{50} M_0$ & $f_{25} M_0$ \\ 
\hline
\hline
\multicolumn{8}{c}{$M_{\rm irr}^{\rm BH}/M_{\rm ADM,0}^{\rm NS}=1$} \\ 
\hline
  3PN  & -0.771 & -2.78 & -0.967 &        &
  1.05(-2) & 2.14(-2) & 3.15(-2) \\ 
  0.15 & 0.155  & -23.6 & 156.   & -401. &
  {\bf 9.80(-3)} & {\it 1.96(-2)} & {\it 2.49(-2)} \\
  0.10 & 1.42   & -128. & 2.45(3)    & -1.66(4) &
  5.48(-3) & {\it 6.81(-3)} & {\it 7.64(-3)} \\
\hline
\multicolumn{8}{c}{$M_{\rm irr}^{\rm BH}/M_{\rm ADM,0}^{\rm NS}=2$} \\ 
\hline
  3PN  & -0.769 & -2.85 & -2.02 &       &
  1.02(-2) & 2.05(-2) & 2.98(-2) \\
  0.15 & 1.66   & -54.6 & 386.   & -1.02(3) &
  {\bf 9.22(-3)} & 1.76(-2) & {\it 2.15(-2) }\\
  0.10 & 1.77   & -105. & 1.43(3)    & -6.94(3) &
  7.48(-3) & {\it 9.72(-3)} & {\it 1.10(-2)}\\
\hline
\multicolumn{8}{c}{$M_{\rm irr}^{\rm BH}/M_{\rm ADM,0}^{\rm NS}=3$} \\ 
\hline
  3PN  & -0.766 & -2.93 & -3.34 &       &
  9.89(-3) & 1.95(-2) & 2.80(-2) \\
  0.15 & 1.28   & -37.0 & 210.  & -494. &
  {\bf 9.57(-3)} & {\bf 1.75(-2)} & 2.28(-2) \\
  0.10 & 1.47   & -74.2 & 815.  & -3.33(3)  &
  {\bf 8.57(-3)} & {\it 1.20(-2)} & {\it 1.38(-2)}\\
\hline
\multicolumn{8}{c}{$M_{\rm irr}^{\rm BH}/M_{\rm ADM,0}^{\rm NS}=5$} \\ 
\hline
  3PN  & -0.762 & -3.05 & -5.20 &        &
  9.51(-3) & 1.84(-2) & 2.60(-2) \\
  0.15 & 0.802  & -26.1 & 138.  & -357. &
  {\bf 1.01(-2)} & {\bf 1.72(-2)} & {\bf 2.21(-2)} \\
  0.10 & 1.03   & -47.7 & 423.  & -1.47(3)  &
  {\bf 9.54(-3)} & 1.45(-2) & {\it 1.71(-2)} \\
\hline
\multicolumn{8}{c}{$M_{\rm irr}^{\rm BH}/M_{\rm ADM,0}^{\rm NS}=10$} \\ 
\hline
  3PN  & -0.757 & -3.18 & -7.36 &        &
  9.14(-3) & 1.73(-2) & 2.42(-2) \\
  0.15 & -0.934 & -14.7 & 122.  & -365. &
  {\bf 6.94(-3)} & {\bf 1.71(-2)} & {\bf 2.30(-2)} \\
  0.10 & 0.305  & -28.5 & 212.  & -643. &
  {\bf 9.35(-3)} & {\bf 1.67(-2)} & {\bf 2.07(-2)} \\
\hline
\end{tabular}
\label{table:dedf}
\end{table}

\section{Comparison with previous results}

We can compare our numerical results with those from previous
attempts to model BHNS binaries.  In particular, we compare with
our previous work that assumed
extreme mass ratios \cite{TanigBFS05}, our comparable-mass results
for a Kerr-Schild background \cite{TanigBFS06}, and the results of
\cite{Grand06} 
for comparable masses and a spatially flat background, as in this study.

\subsection{Comparison with extreme mass ratio results}

\begin{figure}[ht]
\vspace{0.3cm}
\begin{center}
  \includegraphics[width=8cm]{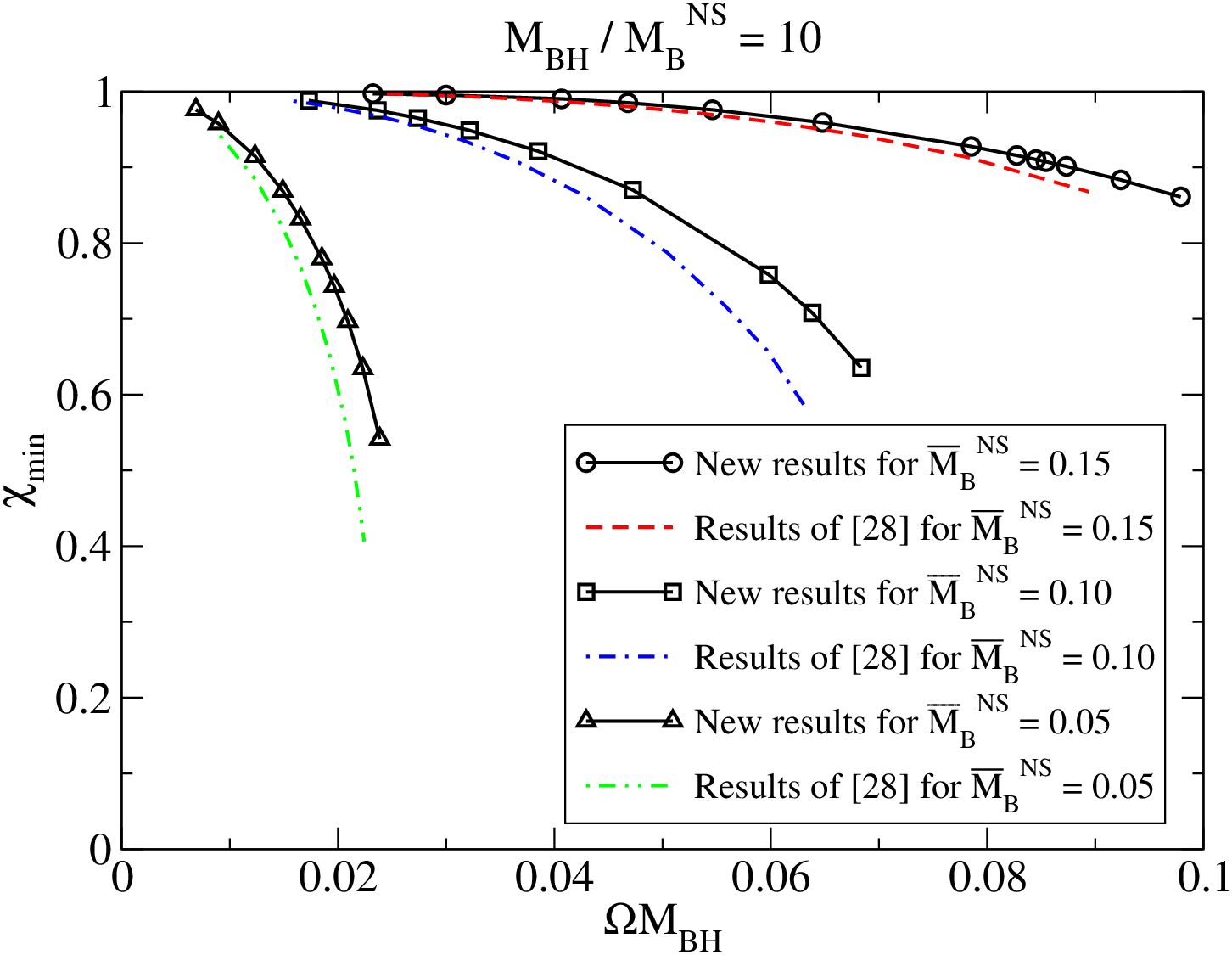}
\end{center}
\caption{Comparison of our new results to those calculated assuming an
  extreme mass ratio for a spatially flat background 
  geometry \cite{TanigBFS05}
  for the minimum of the mass-shedding indicator $\chi_{\rm min}$. Here
  $M_{\rm BH}$ denotes either the black hole irreducible mass (for
  our new results) or the background mass (for the results of
  \cite{TanigBFS05}); see text.}
\label{fig:vsextr}
\end{figure}

In Fig.~\ref{fig:vsextr}, we compare our results for a mass ratio
$M_{\rm irr}^{\rm BH}/M_{\rm B}^{\rm NS}=10$ with those for
$M_{\rm BH}/M_{\rm B}^{\rm NS}=10$ in \cite{TanigBFS05} 
for a spatially flat background, but under the assumption
of an extreme mass ratio.
Since the tidal effects of the neutron star on the black hole are
neglected in the extreme mass-ratio approximation, it is impossible to
evaluate the black hole's irreducible mass. In \cite{TanigBFS05},
therefore, we instead fixed the background mass of the black hole
$M_{\rm BH}$. Since the difference between $M_{\rm irr}^{\rm BH}$ and
$M_{\rm BH}$ scales with the binary's binding energy (see \cite{DenniBP06}
for an analytic, leading-order treatment), the relative difference is
at most a few percent for this mass ratio.

Our present results agree well with those calculated assuming an
extreme mass ratio. The difference in the orbital angular velocity for
the same $\chi_{\rm min}$ is order 10\%. We can explain this 10\%
discrepancy as follows: for a given value of $\chi_{\rm min}$, the 
tidal deformation of the neutron star by the black hole is
approximately the same in the two cases. This implies that the
orbital separation is very similar as well.  However, the orbital
angular velocity in our calculation here is given in terms of the
distance from the center of mass of the binary system to the center of
the neutron star, while for an extreme mass ratio it is given by the
binary separation.  This is because with the extreme mass-ratio assumption,
we assumed that the binary's center of mass is located at the center of
the black hole.
The difference between these two definitions of distance, along with
a change in the black hole mass we use in the definition of the mass
ratio, accounts for the discrepancy in the orbital angular velocities.
To Newtonian order, there is a difference between
$\Omega =\sqrt{M_0/d^3}$ in our calculation
here and $\Omega =\sqrt{M_{\rm BH}/d^3}$ for the extreme mass ratio.
The relative difference between these angular velocities
is about 10\% for the mass ratio in question.

\subsection{Comparison with Kerr-Schild coordinates results}

\begin{figure}[ht]
\vspace{0.3cm}
\begin{center}
  \includegraphics[width=8cm]{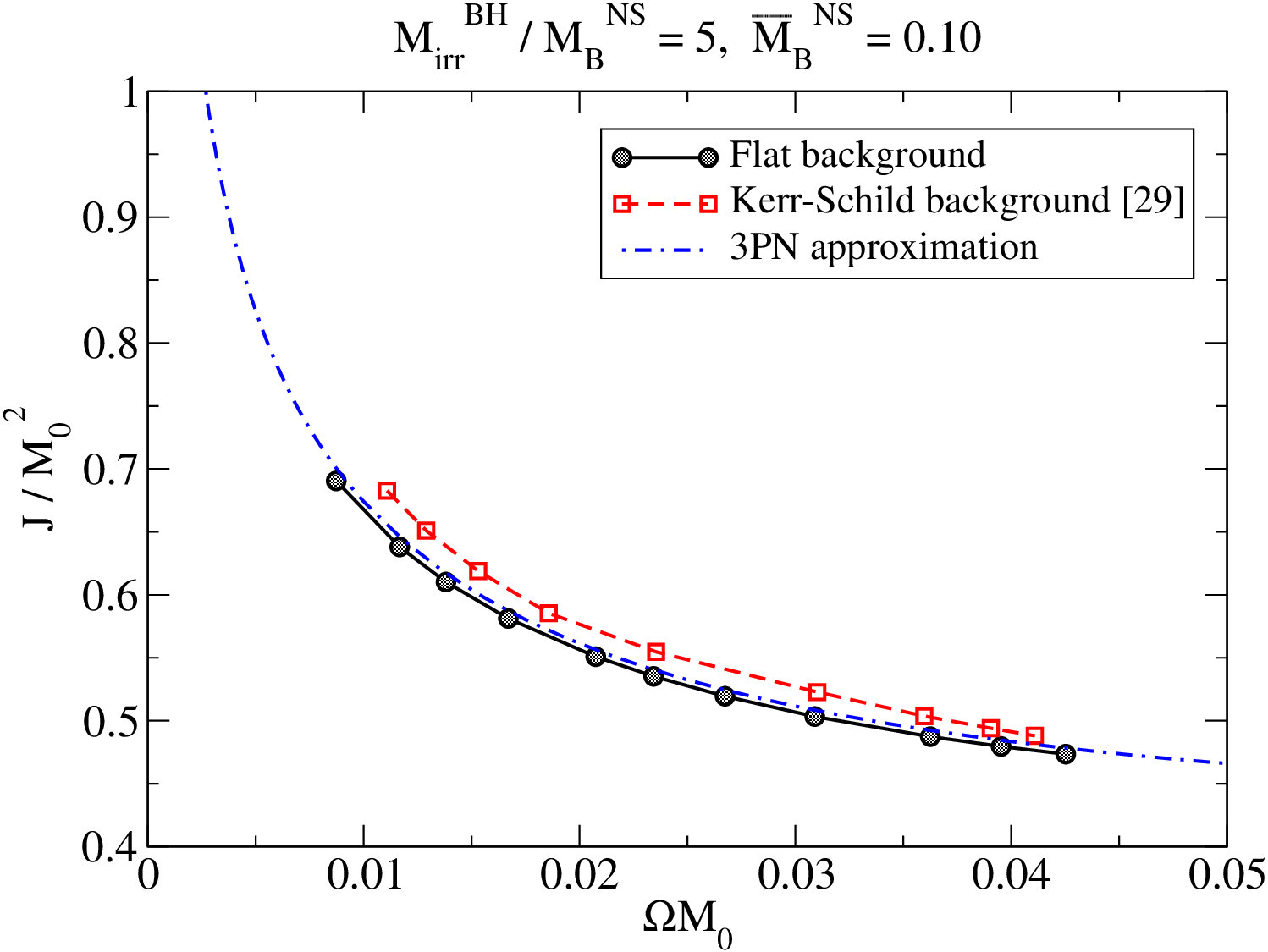}
\end{center}
\caption{Comparison of our new results to those calculated using
  a Kerr-Schild background \cite{TanigBFS06} for
  the total angular momentum.  The dash-dotted line denotes the results
  of the third post-Newtonian approximation \cite{Blanc02}.}
\label{fig:isovsks}
\end{figure}

In Fig. \ref{fig:isovsks}, we compare our new
results for the total
angular momentum with those in \cite{TanigBFS06}, for a mass
ratio $M_{\rm irr}^{\rm BH}/M_{\rm B}^{\rm NS}=5$ and neutron
star mass $\bar{M}_{\rm B}^{\rm NS}=0.10$.  The difference between our
treatment here
and that in \cite{TanigBFS06} is the choice of the
spatial background metric $\tilde{\gamma}_{ij}$.  In this paper, we
assume a flat background, $\tilde{\gamma}_{ij} = \eta_{ij}$, while in
\cite{TanigBFS06} we chose a Kerr-Schild metric.
We see from Fig. \ref{fig:isovsks} that the behavior of the
sequence agrees to within 5\% \cite{angmom} throughout most of 
the sequence.

\subsection{Comparison with previous flat background results}

In \cite{Grand06} (hereafter G06), Grandcl\'ement computed
similar sequences of BHNS binaries, adopting an approach that in some
ways is similar to ours.  Both codes are based on the {\sc Lorene}
spectral method library, but are otherwise completely independent.  In
particular, we impose different inner boundary conditions on the black
hole's excision surface, use different conditions to make the black
hole nonspinning, and adopt a different decomposition of the
equations and variables.  We summarize these differences between our
implementations in Table \ref{table:code}.

\begin{table}[tb]
\caption{Comparison between our implementation and that of G06.  Here
  ``Ex.~BC'' stands for the excision boundary condition, and ``Dec.~of
  (\ref{eq:ham_constr}) -- (\ref{eq:trace_evolv})'' stands for the
  decomposition of the relevant equations and variables.}
\begin{center}
\begin{tabular}{l|c|c} \hline\hline
  & Here & G06 \\ \hline
  $\tilde{\gamma}_{ij}$ & $\eta_{ij}$ &  $\eta_{ij}$ \\
  Ex.~BC: $\psi$, $\beta^i$  & Cook \& Pfeiffer \cite{CookP04} &
  Gourgoulhon  {\em et.al.} \cite{GourgGB02} \\
  Ex.~BC: $\alpha$ & $d \alpha/dr|_{\cal S} =0$ &
  $\alpha|_{\cal S}=0$ \\
  $\Omega_r$ of BH & Cook \& Pfeiffer \cite{CookP04} &
  Caudill {\em et.al.} \cite{CaudiCGP06} \\
  Dec.~of (\ref{eq:ham_constr})-(\ref{eq:trace_evolv}) 
  & See Appendix \ref{appendix:eq} &
  Gourgoulhon  {\em et.al.} \cite{GourgGB02} \\ \hline
\end{tabular}
\end{center}
\label{table:code}
\end{table}

\begin{figure}[ht]
\vspace{0.3cm}
\begin{center}
  \includegraphics[width=8cm]{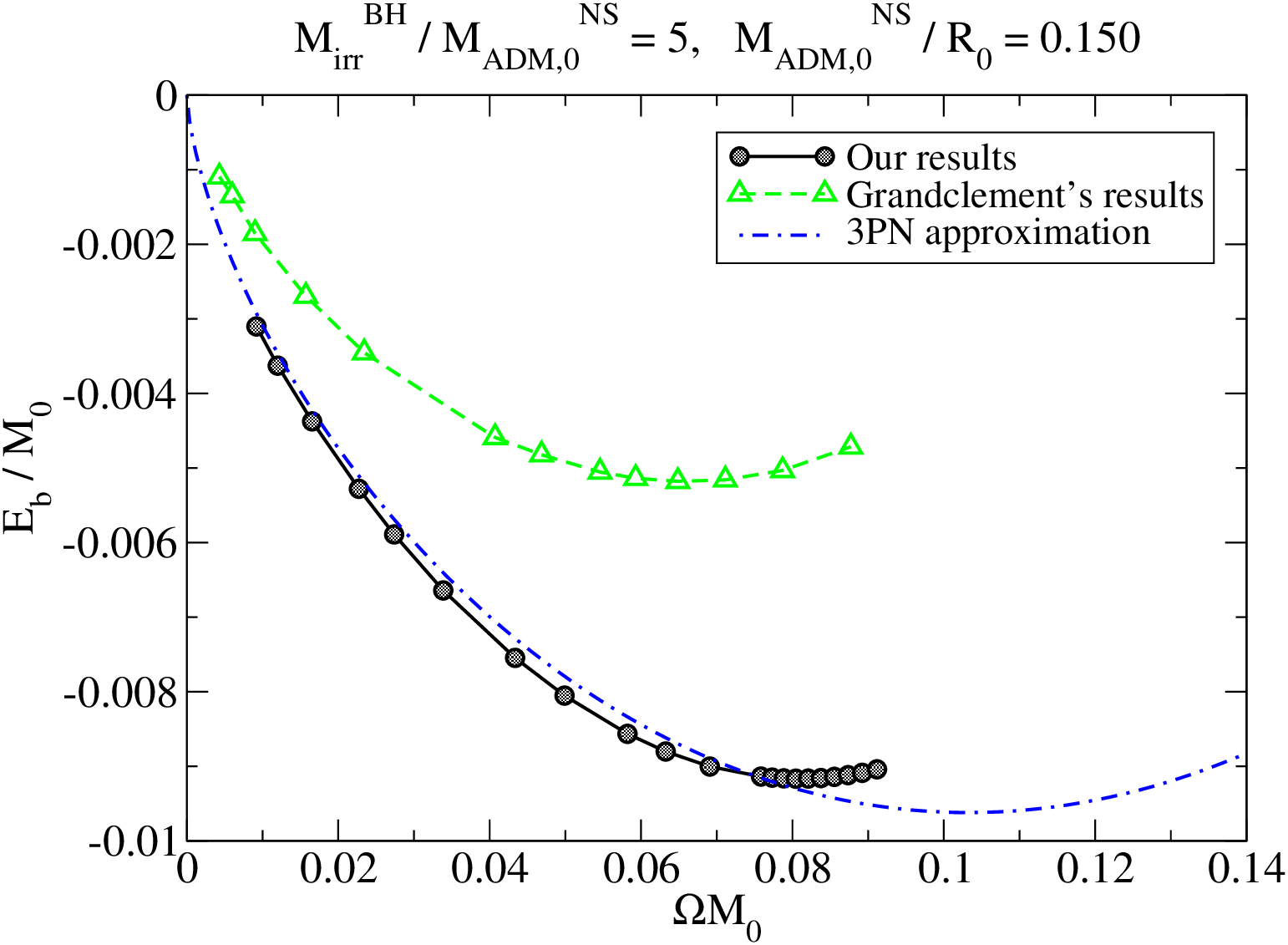}
\end{center}
\caption{Comparison between our results and those of G06 for the
  binding energy as a function of $\Omega M_0$, for a binary of mass
  ratio $M_{\rm irr}^{\rm BH}/M_{\rm ADM,0}^{\rm NS}=5$.  The solid
  line with filled circles marks our results for a neutron star
  compaction of $M_{\rm ADM,0}^{\rm NS}/R_0=0.150$, which corresponds
  to the baryon rest mass of $\bar{M}_{\rm B}^{\rm NS}=0.153$.  The
  dashed line with triangles represents those of G06 for a neutron
  star compaction of 0.150. The dash-dotted line denotes the results
  of the third post-Newtonian approximation \cite{Blanc02}.}
\label{fig:bind_compare}
\end{figure}

To compare our results with those of G06, we show in
Fig.~\ref{fig:bind_compare} the binding energy as a function of
$\Omega M_0$ for two sequences with the same physical parameters.
Both sequences are for a binary with mass ratio $M_{\rm irr}^{\rm
BH}/M_{\rm ADM,0}^{\rm NS}=5$.  Our results are for a neutron star
rest mass of $\bar{M}_{\rm B}^{\rm NS}=0.153$, which corresponds to a
compaction of $M_{\rm ADM,0}^{\rm NS}/R_0=0.150$.  We find agreement
with the third-order post-Newtonian result to within 3\% except for
configurations inside the turning point, and hence inside the ISCO.  
We also include
the binding energy for a neutron star compaction of 0.150 as computed
by G06 (and as found in Fig.~3 of G06).  Clearly this curve shows a
much larger deviation from both our results and the post-Newtonian
ones, by about a factor of two.  We speculate that this deviation
could be caused by the larger numerical error in the calculations of
G06.  According to G06, the relative virial error is as large as 2\%.
If the masses themselves carry an error of 2\%, the binding energy may
have an error as large as 100\% or more, as we have outlined in the
paragraph including Eqs.~(\ref{eq:errorADM}) and (\ref{eq:error}).
This is consistent with the deviations found in
Fig.~\ref{fig:bind_compare}.

\section{Summary}

We construct quasiequilibrium sequences of BHNS binaries in general
relativity.  We solve the constraint equations of general relativity,
decomposed in the conformal thin-sandwich formalism and
subject to the black hole boundary conditions of \cite{CookP04},
together with the relativistic equations of hydrostatic equilibrium.
In contrast to our earlier approach in \cite{TanigBFS06},
where we solved these equations for a Kerr-Schild background, we now
adopt a spatially flat background.  We also employ a new decomposition
of the equations and variables that leads to a significant improvement
in the numerical accuracy.

We construct constant-mass sequences of mass ratios $M_{\rm
irr}^{\rm BH}/M_{\rm ADM,0}^{\rm NS}=1$, 2, 3, 5, and 10 for neutron
star baryon rest masses of $\bar{M}_{\rm B}^{\rm NS}=0.15$ and 0.10.
We find excellent agreement between our calculated binding energies
and the third-order post-Newtonian approximation, especially for large
binary separations, but also agreeing to within 5\% at small
separations.  We locate the onset of tidal disruption and, for large
mass ratios and neutron star compactions, the ISCO.

We find some evidence that our ``leading-order approximation'' for
constructing nonspinning black holes leads to small errors,
and plan to implement an improved condition in the near future.

\section*{ACKNOWLEDGMENTS}

J.A.F.~gratefully acknowledges support through NSF Grant
AST-0401533. This paper was supported in part by NSF Grants
PHY-0205155 and PHY-0345151 as well as NASA Grant NNG04GK54G to
University of Illinois at Urbana-Champaign, and NSF Grant PHY-0456917
to Bowdoin College.

\appendix

\section{Decomposition of the gravitational field variables}
\label{appendix:eq}

For the construction of binary configurations with codes that are
based on spherical polar coordinates it is natural to use two computational
domains, each one centered on
one of the binary companions.  An equation of the form
\begin{equation}
  \underline{\Delta} \phi = \mbox{RHS}
\end{equation}
can then be split into two equations
\begin{eqnarray} \label{split}
  \underline{\Delta} \phi_{\rm BH} &=& \mbox{RHS}_{\rm BH}
  \label{split1} \\
  \underline{\Delta} \phi_{\rm NS} &=& \mbox{RHS}_{\rm NS},
  \label{split2}
\end{eqnarray}
where $\phi = \phi_{\rm BH} + \phi_{\rm NS}$ and
$\mbox{RHS} = \mbox{RHS}_{\rm BH} + \mbox{RHS}_{\rm NS}$.
The two equations (\ref{split1}) and (\ref{split2}) can then be solved
on two computational domains, one centered on the black hole and the
other centered on the neutron star.

Clearly, the separation of the source terms 
$\mbox{RHS}$ into its two parts $\mbox{RHS}_{\rm BH}$ and 
$\mbox{RHS}_{\rm NS}$ is not unique.  One guiding principle is to move
those parts of $\mbox{RHS}$ that are large in the neighborhood of the
black hole into $\mbox{RHS}_{\rm BH}$, and likewise for the neutron
star. Another principle is that each of the source terms should
asymptotically coincide with that for the isolated black hole and
neutron star when the orbital separation is large. In addition, we use
this decomposition freedom to deal with the fact that
metric quantities are not defined inside the excised parts of the
black hole.  It is not clear how to treat such functions if they
appear in $\mbox{RHS}_{\rm NS}$.  One approach is to artificially fill
the resulting ``hole'' in the function with a smooth function (see,
e.g.~\cite{GourgGB02}). Instead, we carefully
separate the source terms in such a way that no metric quantities that
are excised in the black hole interior affect the right-hand side of
the neutron star equation.

Specifically, we first decompose the metric quantities as
\begin{eqnarray}
  &&\alpha = \alpha_{\rm BH} +\alpha_{\rm NS}, \\
  &&\beta^i = \beta^i_{\rm BH} +\beta^i_{\rm NS}
  +\beta^i_{\rm rot}, \\
  &&\psi = \psi_{\rm BH} +\psi_{\rm NS},
\end{eqnarray}
where $\beta^i_{\rm rot}$ is the rotational shift vector.
We then decompose the gravitational field equations as follows.
Equation (\ref{eq:ham_constr}) becomes
\begin{eqnarray}
  &&\underline{\Delta} \psi_{\rm BH} = -{1 \over 8} \psi^{-7}
  \tilde{A}_{ij} \tilde{A}^{ij} \nonumber \\
  &&\hspace{20pt}+{1 \over 8} (\psi_{\rm NS}+c_{\psi})^{-7}
  \tilde{A}_{ij}^{\rm NS} \tilde{A}^{ij}_{\rm NS}, \\
  &&\underline{\Delta} \psi_{\rm NS} = -2 \pi \psi^5 \rho
  -{1 \over 8} (\psi_{\rm NS}+c_{\psi})^{-7}
  \tilde{A}_{ij}^{\rm NS} \tilde{A}^{ij}_{\rm NS}, \label{app:ham_ns}
\end{eqnarray}
Eq.~(\ref{eq:mom_constr}) is decomposed as
\begin{eqnarray}
  &&\underline{\Delta} \beta^i_{\rm BH} +{1 \over 3} \partial^i
  (\partial_j \beta^j_{\rm BH}) =2 \tilde{A}^{ij} \partial_j
  (\alpha \psi^{-6}) \nonumber \\
  &&\hspace{20pt}-2 \tilde{A}^{ij}_{\rm NS} \partial_j
  [(\alpha_{\rm NS}+c_{\alpha})
    (\psi_{\rm NS}+c_{\psi})^{-6}], \\
  &&\underline{\Delta} \beta^i_{\rm NS} +{1 \over 3} \partial^i
  (\partial_j \beta^j_{\rm NS}) =16 \pi \alpha \psi^4 j^i
  \nonumber \\
  &&\hspace{20pt}+2 \tilde{A}^{ij}_{\rm NS} \partial_j
  [(\alpha_{\rm NS}+c_{\alpha})
    (\psi_{\rm NS}+c_{\psi})^{-6}], \label{app:mom_ns}
\end{eqnarray}
and Eq.~(\ref{eq:trace_evolv}) is written as
\begin{eqnarray}
  &&\underline{\Delta} \alpha_{\rm BH} =\alpha \psi^{-8}
  \tilde{A}_{ij} \tilde{A}^{ij} -2 \eta^{ij} \partial_i \alpha
  \partial_j \ln \psi \nonumber \\
  &&\hspace{20pt}-(\alpha_{\rm NS}+c_{\alpha})
  (\psi_{\rm NS}+c_{\psi})^{-8}
  \tilde{A}_{ij}^{\rm NS} \tilde{A}^{ij}_{\rm NS} \nonumber \\
  &&\hspace{20pt}+2 \eta^{ij} \partial_i (\alpha_{\rm NS}+c_{\alpha})
  \partial_j \ln (\psi_{\rm NS}+c_{\psi}), \\
  &&\underline{\Delta} \alpha_{\rm NS} =4 \pi \alpha \psi^4
  (\rho +S) \nonumber \\
  &&\hspace{20pt}+(\alpha_{\rm NS}+c_{\alpha})
  (\psi_{\rm NS}+c_{\psi})^{-8}
  \tilde{A}_{ij}^{\rm NS} \tilde{A}^{ij}_{\rm NS} \nonumber \\
  &&\hspace{20pt}-2 \eta^{ij} \partial_i (\alpha_{\rm NS}+c_{\alpha})
  \partial_j \ln (\psi_{\rm NS}+c_{\psi}), \label{app:trace_ns}
\end{eqnarray}
where we define $\bar{A}^{ij}_{\rm NS}$ as
\begin{equation}
  \tilde{A}^{ij}_{\rm NS} \equiv {(\psi_{\rm NS}+c_{\psi})^6
    \over 2 (\alpha_{\rm NS}+c_{\alpha})} \Bigl( \partial^i
  \beta^j_{\rm NS} +\partial^j \beta^i_{\rm NS}
  -{2 \over 3} \eta^{ij} \partial_k \beta^k_{\rm NS} \Bigr).
\end{equation}

Note that the neutron star equations 
(\ref{app:ham_ns}), (\ref{app:mom_ns}), and (\ref{app:trace_ns})
contain metric quantities like $\alpha$ and $\psi$ that are not defined
inside the excised black hole, but only in products with matter quantities
that vanish outside the neutron star.  As a consequence, the neutron
star equations are regular everywhere.  All singular, or un-defined
terms have been moved to the black hole equations, where they
are harmless since they are excised from the computational grid.

We finally point out that the outer boundary conditions of the total
lapse function and total conformal factor are asymptotically flat,
i.e., $\alpha|_{r \rightarrow \infty}=1$ and
$\psi|_{r \rightarrow \infty}=1$. This implies that each part of the
metric quantities, i.e., $\alpha_{\rm BH}$,
$\alpha_{\rm NS}$, $\psi_{\rm BH}$, and
$\psi_{\rm NS}$, cannot go to unity individually at infinity.
We set the outer boundary conditions for these quantities as
$\alpha_{\rm BH}|_{r \rightarrow \infty}=0.5$,
$\alpha_{\rm NS}|_{r \rightarrow \infty}=0.5$,
$\psi_{\rm BH}|_{r \rightarrow \infty}=0.5$, and
$\psi_{\rm NS}|_{r \rightarrow \infty}=0.5$
for the convenience of the computation.
We therefore insert constants $c_{\alpha}=0.5$ and $c_{\psi}=0.5$ to
ensure that the total quantities take on their proper asymptotic values.

\section{Tables of sequences}
\label{appendix:table}

We summarize our results in Tables \ref{table:seqM015} and
\ref{table:seqM01}. In these tables, we tabulate the orbital angular
velocity $\Omega$, binding energy $E_{\rm b}$, total angular momentum
$J$, decrease in the maximum density parameter $\delta q_{\rm max}$,
minimum of the mass-shedding indicator $\chi_{\rm min}$,
and fractional difference $\delta M$ between the ADM mass and the
Komar mass along a sequence.

\begin{table*}[ht]
\caption{Physical parameters for a binary sequence with neutron star
  baryon rest mass $\bar{M}_{\rm B}^{\rm NS}=0.15$. The ADM mass, and
  the isotropic coordinate radius of the neutron star in isolation are
  $\bar{M}_{\rm ADM,0}^{\rm NS}=0.139$ and $\bar{r}_0=0.815$
  ($\kappa=1$). The neutron star compaction is
  $M_{\rm ADM,0}^{\rm NS}/R_0=0.145$ where $R_0$ is the areal radius of
  the isolated neutron star. We list the binding energy $E_{\rm b}$,
  total angular momentum $J$, orbital angular velocity $\Omega$,
  decrease of the maximum density parameter $\delta q_{\rm max}$,
  minimum of the mass-shedding indicator $\chi_{\rm min}$, and
  fractional difference $\delta M$ between the ADM mass and the Komar
  mass. The prefix $\dagger$ denotes the position of the turning
  point.}
\begin{center}
\begin{tabular}{rcccccc} \hline\hline
  \multicolumn{7}{c}{Mass ratio:
    $M_{\rm irr}^{\rm BH}/M_{\rm ADM,0}^{\rm NS}=1$} \\
  $d/M_0$&$\Omega M_0$&$E_{\rm b}/M_0$&$J/M_0^2$&
  $\delta q_{\rm max}$&$\chi_{\rm min}$&$\delta M$ \\ \hline
  15.28 & 0.0153 & -7.34(-3) & 1.107 & -9.56(-5) & 0.961 & 6.66(-6) \\
  13.10 & 0.0191 & -8.36(-3) & 1.048 & -4.16(-4) & 0.934 & 2.83(-5) \\
  10.91 & 0.0246 & -9.71(-3) & 0.985 & -1.45(-3) & 0.871 & 1.95(-5) \\
   9.82 & 0.0285 & -1.06(-2) & 0.952 & -2.87(-3) & 0.803 & 5.10(-6) \\
   9.17 & 0.0313 & -1.11(-2) & 0.933 & -4.56(-3) & 0.731 & 9.44(-6) \\
   8.73 & 0.0335 & -1.15(-2) & 0.920 & -6.43(-3) & 0.648 & 2.73(-5) \\
   8.51 & 0.0347 & -1.18(-2) & 0.914 & -7.80(-3) & 0.575 & 4.76(-5) \\
   8.30 & 0.0359 & -1.20(-2) & 0.907 & -1.03(-2) & 0.435 & 1.42(-4) \\
   \hline
  \multicolumn{7}{c}{Mass ratio:
    $M_{\rm irr}^{\rm BH}/M_{\rm ADM,0}^{\rm NS}=2$} \\
  $d/M_0$&$\Omega M_0$&$E_{\rm b}/M_0$&$J/M_0^2$&
  $\delta q_{\rm max}$&$\chi_{\rm min}$&$\delta M$ \\ \hline
  20.37 & 0.0101 & -5.15(-3) & 1.089 & -3.65(-5) & 0.992 & 4.45(-4) \\
  11.64 & 0.0224 & -8.30(-3) & 0.891 & -1.19(-3) & 0.952 & 1.33(-4) \\
   8.73 & 0.0334 & -1.04(-2) & 0.815 & -3.49(-3) & 0.870 & 3.13(-5) \\
   8.00 & 0.0376 & -1.10(-2) & 0.796 & -5.06(-3) & 0.822 & 2.39(-5) \\
   7.28 & 0.0427 & -1.17(-2) & 0.777 & -7.99(-3) & 0.742 & 3.70(-5) \\
   6.84 & 0.0464 & -1.22(-2) & 0.765 & -1.12(-2) & 0.661 & 6.78(-5) \\
   6.55 & 0.0491 & -1.26(-2) & 0.758 & -1.46(-2) & 0.566 & 1.14(-4) \\
   6.40 & 0.0506 & -1.27(-2) & 0.754 & -1.70(-2) & 0.472 & 1.80(-4) \\
   \hline
  \multicolumn{7}{c}{Mass ratio:
    $M_{\rm irr}^{\rm BH}/M_{\rm ADM,0}^{\rm NS}=3$} \\
  $d/M_0$&$\Omega M_0$&$E_{\rm b}/M_0$&$J/M_0^2$&
  $\delta q_{\rm max}$&$\chi_{\rm min}$&$\delta M$ \\ \hline
  19.65 & 0.0105 & -4.50(-3) & 0.900 & -1.01(-4) & 0.995 & 6.33(-4) \\
  13.10 & 0.0189 & -6.41(-3) & 0.778 & -1.15(-3) & 0.981 & 4.37(-4) \\
   9.82 & 0.0282 & -8.09(-3) & 0.709 & -2.79(-3) & 0.952 & 2.87(-4) \\
   7.64 & 0.0398 & -9.70(-3) & 0.661 & -5.86(-3) & 0.888 & 1.71(-4) \\
   6.55 & 0.0489 & -1.07(-2) & 0.638 & -9.61(-3) & 0.808 & 1.31(-4) \\
   6.01 & 0.0548 & -1.12(-2) & 0.626 & -1.34(-2) & 0.735 & 1.36(-4) \\
   5.46 & 0.0620 & -1.17(-2) & 0.616 & -2.08(-2) & 0.597 & 1.98(-4) \\
   5.35 & 0.0636 & -1.18(-2) & 0.614 & -2.31(-2) & 0.543 & 2.32(-4) \\
   \hline
  \multicolumn{7}{c}{Mass ratio:
    $M_{\rm irr}^{\rm BH}/M_{\rm ADM,0}^{\rm NS}=5$} \\
  $d/M_0$&$\Omega M_0$&$E_{\rm b}/M_0$&$J/M_0^2$&
  $\delta q_{\rm max}$&$\chi_{\rm min}$&$\delta M$ \\ \hline
  18.19 & 0.0117 & -3.56(-3) & 0.642 & -3.87(-4) & 0.997 & 7.20(-4) \\
  11.64 & 0.0221 & -5.19(-3) & 0.550 & -2.12(-3) & 0.988 & 6.02(-4) \\
   7.28 & 0.0422 & -7.44(-3) & 0.481 & -7.28(-3) & 0.945 & 4.12(-4) \\
   5.46 & 0.0617 & -8.71(-3) & 0.454 & -1.43(-2) & 0.856 & 2.86(-4) \\
   4.59 & 0.0769 & -9.12(-3) & 0.443 & -2.15(-2) & 0.736 & 2.26(-4) \\
   $\dagger$
   4.37 & 0.0817 & -9.14(-3) & 0.442 & -2.49(-2) & 0.682 & 2.22(-4) \\
   4.23 & 0.0852 & -9.13(-3) & 0.441 & -2.81(-2) & 0.635 & 2.27(-4) \\
   4.08 & 0.0889 & -9.08(-3) & 0.440 & -3.23(-2) & 0.566 & 2.36(-4) \\
   \hline
  \multicolumn{7}{c}{Mass ratio:
    $M_{\rm irr}^{\rm BH}/M_{\rm ADM,0}^{\rm NS}=10$} \\
  $d/M_0$&$\Omega M_0$&$E_{\rm b}/M_0$&$J/M_0^2$&
  $\delta q_{\rm max}$&$\chi_{\rm min}$&$\delta M$ \\ \hline
  11.91 & 0.0212 & -2.82(-3) & 0.326 & -2.39(-3) & 0.997 & 6.21(-4) \\
   7.94 & 0.0372 & -3.80(-3) & 0.291 & -5.21(-3) & 0.991 & 3.81(-4) \\
   6.35 & 0.0501 & -4.39(-3) & 0.276 & -7.14(-3) & 0.978 & 2.92(-4) \\
   5.56 & 0.0596 & -4.73(-3) & 0.269 & -8.57(-3) & 0.963 & 2.69(-4) \\
   4.57 & 0.0764 & -5.10(-3) & 0.261 & -1.19(-2) & 0.925 & 2.80(-4) \\
   $\dagger$
   4.25 & 0.0835 & -5.14(-3) & 0.259 & -1.40(-2) & 0.904 & 2.96(-4) \\
   3.97 & 0.0906 & -5.10(-3) & 0.258 & -1.67(-2) & 0.879 & 3.18(-4) \\
   3.58 & 0.1028 & -4.82(-3) & 0.257 & -2.37(-2) & 0.823 & 3.72(-4) \\
   \hline
\end{tabular}
\end{center}
\label{table:seqM015}
\end{table*}

\begin{table*}[ht]
\caption{Same as Table \ref{table:seqM015} but for the neutron-star
  baryon rest mass $\bar{M}_{\rm B}^{\rm NS}=0.10$. The ADM mass, and
  the isotropic coordinate radius of the neutron star in isolation are
  $\bar{M}_{\rm ADM,0}^{\rm NS}=0.0956$ and $\bar{r}_0=0.990$
  ($\kappa=1$). The neutron star compaction is
  $M_{\rm ADM,0}^{\rm NS}/R_0=0.0879$.}
\begin{center}
\begin{tabular}{rcccccc} \hline\hline
  \multicolumn{7}{c}{Mass ratio:
    $M_{\rm irr}^{\rm BH}/M_{\rm ADM,0}^{\rm NS}=1$} \\
  $d/M_0$&$\Omega M_0$&$E_{\rm b}/M_0$&$J/M_0^2$&
  $\delta q_{\rm max}$&$\chi_{\rm min}$&$\delta M$ \\ \hline
  31.84 & 0.00533 & -3.74(-3) & 1.498 & -3.84(-5) & 0.976 & 4.38(-6) \\
  25.47 & 0.00738 & -4.59(-3) & 1.362 & -2.00(-4) & 0.949 & 1.44(-5) \\
  20.70 & 0.00997 & -5.55(-3) & 1.250 & -7.50(-4) & 0.894 & 8.77(-6) \\
  17.51 & 0.01267 & -6.44(-3) & 1.171 & -2.21(-3) & 0.797 & 1.38(-6) \\
  15.92 & 0.01453 & -7.00(-3) & 1.130 & -4.31(-3) & 0.683 & 1.03(-5) \\
  15.28 & 0.01541 & -7.24(-3) & 1.114 & -5.87(-3) & 0.596 & 1.77(-5) \\
  14.97 & 0.01588 & -7.37(-3) & 1.106 & -6.98(-3) & 0.530 & 2.41(-5) \\
  14.81 & 0.01613 & -7.43(-3) & 1.102 & -7.68(-3) & 0.486 & 2.89(-5) \\
  \hline
  \multicolumn{7}{c}{Mass ratio:
    $M_{\rm irr}^{\rm BH}/M_{\rm ADM,0}^{\rm NS}=2$} \\
  $d/M_0$&$\Omega M_0$&$E_{\rm b}/M_0$&$J/M_0^2$&
  $\delta q_{\rm max}$&$\chi_{\rm min}$&$\delta M$ \\ \hline
  33.96 & 0.00483 & -3.15(-3) & 1.364 & -5.89(-5) & 0.990 & 1.19(-4) \\
  21.23 & 0.00958 & -4.85(-3) & 1.121 & -4.27(-4) & 0.952 & 2.48(-5) \\
  16.98 & 0.01321 & -5.91(-3) & 1.027 & -1.18(-3) & 0.899 & 1.50(-6) \\
  13.80 & 0.01774 & -7.07(-3) & 0.951 & -3.57(-3) & 0.785 & 2.83(-6) \\
  12.74 & 0.01986 & -7.55(-3) & 0.925 & -5.82(-3) & 0.699 & 1.62(-5) \\
  12.10 & 0.02134 & -7.88(-3) & 0.910 & -8.20(-3) & 0.614 & 3.27(-5) \\
  11.68 & 0.02244 & -8.10(-3) & 0.899 & -1.06(-2) & 0.523 & 5.11(-5) \\
  11.46 & 0.02303 & -8.22(-3) & 0.894 & -1.23(-2) & 0.454 & 6.41(-5) \\
  \hline
  \multicolumn{7}{c}{Mass ratio:
    $M_{\rm irr}^{\rm BH}/M_{\rm ADM,0}^{\rm NS}=3$} \\
  $d/M_0$&$\Omega M_0$&$E_{\rm b}/M_0$&$J/M_0^2$&
  $\delta q_{\rm max}$&$\chi_{\rm min}$&$\delta M$ \\ \hline
  19.10 & 0.0111 & -4.53(-3) & 0.905 & -6.79(-4) & 0.961 & 8.29(-5) \\
  15.92 & 0.0144 & -5.31(-3) & 0.844 & -1.27(-3) & 0.929 & 5.01(-5) \\
  12.74 & 0.0198 & -6.41(-3) & 0.779 & -3.13(-3) & 0.849 & 2.51(-5) \\
  11.94 & 0.0217 & -6.76(-3) & 0.762 & -4.22(-3) & 0.810 & 2.41(-5) \\
  11.14 & 0.0239 & -7.14(-3) & 0.745 & -5.97(-3) & 0.753 & 2.86(-5) \\
  10.35 & 0.0264 & -7.57(-3) & 0.728 & -9.05(-3) & 0.665 & 4.46(-5) \\
  10.03 & 0.0276 & -7.75(-3) & 0.721 & -1.10(-2) & 0.611 & 5.70(-5) \\
   9.55 & 0.0295 & -8.04(-3) & 0.711 & -1.52(-2) & 0.484 & 8.75(-5) \\
  \hline
  \multicolumn{7}{c}{Mass ratio:
    $M_{\rm irr}^{\rm BH}/M_{\rm ADM,0}^{\rm NS}=5$} \\
  $d/M_0$&$\Omega M_0$&$E_{\rm b}/M_0$&$J/M_0^2$&
  $\delta q_{\rm max}$&$\chi_{\rm min}$&$\delta M$ \\ \hline
  16.98 & 0.0131 & -3.73(-3) & 0.638 & -1.15(-3) & 0.975 & 1.64(-4) \\
  12.74 & 0.0197 & -4.79(-3) & 0.575 & -2.57(-3) & 0.936 & 1.20(-4) \\
  10.61 & 0.0254 & -5.55(-3) & 0.542 & -4.54(-3) & 0.884 & 9.89(-5) \\
   9.55 & 0.0294 & -6.02(-3) & 0.524 & -6.55(-3) & 0.834 & 9.23(-5) \\
   8.49 & 0.0345 & -6.56(-3) & 0.507 & -1.04(-2) & 0.748 & 9.67(-5) \\
   7.96 & 0.0376 & -6.85(-3) & 0.499 & -1.40(-2) & 0.676 & 1.11(-4) \\
   7.43 & 0.0413 & -7.17(-3) & 0.491 & -2.00(-2) & 0.556 & 1.41(-4) \\
   7.33 & 0.0421 & -7.23(-3) & 0.489 & -2.17(-2) & 0.518 & 1.49(-4) \\
  \hline
  \multicolumn{7}{c}{Mass ratio:
    $M_{\rm irr}^{\rm BH}/M_{\rm ADM,0}^{\rm NS}=10$} \\
  $d/M_0$&$\Omega M_0$&$E_{\rm b}/M_0$&$J/M_0^2$&
  $\delta q_{\rm max}$&$\chi_{\rm min}$&$\delta M$ \\ \hline
  11.58 & 0.0224 & -3.02(-3) & 0.330 & -3.52(-3) & 0.976 & 1.58(-4) \\
   8.11 & 0.0365 & -4.00(-3) & 0.297 & -8.37(-3) & 0.925 & 1.25(-4) \\
   6.95 & 0.0448 & -4.43(-3) & 0.286 & -1.27(-2) & 0.876 & 1.16(-4) \\
   5.79 & 0.0568 & -4.88(-3) & 0.276 & -2.22(-2) & 0.770 & 1.24(-4) \\
   5.50 & 0.0607 & -4.98(-3) & 0.273 & -2.67(-2) & 0.723 & 1.34(-4) \\
   5.21 & 0.0650 & -5.07(-3) & 0.271 & -3.30(-2) & 0.657 & 1.45(-4) \\
   4.98 & 0.0688 & -5.13(-3) & 0.270 & -3.99(-2) & 0.579 & 1.54(-4) \\
   4.93 & 0.0698 & -5.14(-3) & 0.269 & -4.16(-2) & 0.553 & 1.55(-4) \\
   \hline
\end{tabular}
\end{center}
\label{table:seqM01}
\end{table*}



\begin{thebibliography}{99}

\bibitem{LIGO}
S. J. Waldman (LIGO Scientific Collaboration),
Class. Quantum Grav. {\bf 23}, S653 (2006).

\bibitem{GEO}
H. L\"uck (GEO600 Collaboration),
Class. Quantum Grav. {\bf 23}, S71 (2006).

\bibitem{TAMA}
M. Ando (TAMA Collabaration),
Class. Quantum Grav. {\bf 22}, S881 (2005).

\bibitem{VIRGO}
F. Acernese (VIRGO Collaboration),
Class. Quantum Grav. {\bf 23}, S635 (2006).

\bibitem{Barge06}
E. Barger,
in {\it Proceedings of the 16th Annual Astrophysics Conference on
  Gamma Ray Bursts in the Swift Era},
Maryland, edited by S. Holt, N. Gehrels, and J. Nousek,
astro-ph/0602004.

\bibitem{ShibaT06}
M. Shibata and K. Taniguchi,
Phys. Rev. D {\bf 73}, 064027 (2006).

\bibitem{FaberBST06}
J. A. Faber, T. W. Baumgarte, S. L. Shapiro, and K. Taniguchi,
Astrophys. J. {\bf 641}, L93 (2006).

\bibitem{PriceR06}
D. J. Price and S. Rosswog,
Science {\bf 312}, 719 (2006).

\bibitem{OechsJ06}
R. Oechslin and H.-T. Janka,
in {\it Proceedings of the Albert Einstein Century International
Conference},
Paris, France 2005, edited by J.-M. Alimi and A. Fuzfa,
astro-ph/0604562.

\bibitem{Chand69}
S. Chandrasekhar,
{\it Ellipsoidal Figures of Equilibrium}
(Yale University Press, New Heaven, CT, 1969).

\bibitem{Fishb73}
L. G. Fishbone,
Astrophys. J. {\bf 185}, 43 (1973).

\bibitem{LaiRS93}
D. Lai, F, A, Rasio, and S. L. Shapiro,
Astrophys. J. Suppl. Ser. {\bf 88}, 205 (1993).

\bibitem{LaiW96}
D. Lai and A. G. Wiseman,
Phys. Rev. D {\bf 54}, 3958 (1996).

\bibitem{TanigN96}
K. Taniguchi and T. Nakamura,
Prog. Theor. Phys. {\bf 96}, 693 (1996).

\bibitem{Shiba96}
M. Shibata,
Prog. Theor. Phys. {\bf 96}, 917 (1996).

\bibitem{UryuE99}
K. Ury\=u and Y. Eriguchi,
Mon. Not. R. Astron. Soc. {\bf 303}, 329 (1999).

\bibitem{WiggiL00}
P. Wiggins and D. Lai,
Astrophys. J. {\bf 532}, 530 (2000).

\bibitem{IshiiSM05}
M. Ishii, M. Shibata, and Y. Mino,
Phys. Rev. D {\bf 71}, 044017 (2005).

\bibitem{Mashh75}
B. Mashhoon,
Astrophys. J. {\bf 197}, 705 (1975).

\bibitem{CarteL83}
B. Carter and J.-P. Luminet,
Astron. Astrophys. {\bf 121}, 97 (1983);
Mon. Not. R. Astron. Soc. {\bf 212}, 23 (1985).

\bibitem{Marck83}
J.-A. Marck,
Proc. R. Soc. London A {\bf 385}, 431 (1983).

\bibitem{LeeK99}
W. H. Lee and W. Klu\'zniak,
Astrophys. J. {\bf 526}, 178 (1999);
Mon. Not. R. Astron. Soc. {\bf 308}, 780 (1999).

\bibitem{Lee00}
W. H. Lee,
Mon. Not. R. Astron. Soc. {\bf 318}, 606 (2000);
{\bf 328}, 583 (2001).

\bibitem{RosswSW04}
S. Rosswog, R. Speith, and G. A. Wynn,
Mon. Not. R. Astron. Soc. {\bf 351}, 1121 (2004).

\bibitem{KobayLPM04}
S. Kobayashi, P. Laguna, E. S. Phinney, and P. M\'esz\'aros,
Astrophys. J. {\bf 615}, 855 (2004).

\bibitem{Mille01}
M. Miller, gr-qc/0106017.

\bibitem{BaumgSS04}
T. W. Baumgarte, M. L. Skoge, and S. L. Shapiro,
Phys. Rev. D {\bf 70}, 064040 (2004).

\bibitem{TanigBFS05}
K. Taniguchi, T. W. Baumgarte, J. A. Faber, and S. L. Shapiro,
Phys. Rev. D {\bf 72}, 044008 (2005).

\bibitem{TanigBFS06}
K. Taniguchi, T. W. Baumgarte, J. A. Faber, and S. L. Shapiro,
Phys. Rev. D {\bf 74}, 041502(R) (2006).

\bibitem{Grand06}
P. Grandcl\'ement, Phys. Rev. D. {\bf 74}, 124002 (2006).

\bibitem{FaberBSTR06}
J. A. Faber, T. W. Baumgarte, S. L. Shapiro, K. Taniguchi, and
F. A. Rasio,
Phys. Rev. D {\bf 73}, 024012 (2006).

\bibitem{SopueSL06}
C. F. Sopuerta, U. Sperhake, and P. Laguna,
Class. Quantum Grav. {\bf 23}, S579 (2006).

\bibitem{LofflRA06}
F. L\"offler, L. Rezzolla, and M. Ansorg,
Phys. Rev. D. {\bf 74}, 104018 (2006).

\bibitem{ShibaU06}
M. Shibata and K. Ury\=u,
Phys. Rev. D. {\bf 74}, 121503(R) (2006).

\bibitem{Valli00}
M. Vallisneri, Phys. Rev. Lett. {\bf 84}, 3519 (2000)

\bibitem{York99}
J. W. York, Jr, Phys. Rev. Lett. {\bf 82}, 1350 (1999).
                                                                              
\bibitem{Cook00}
G. B. Cook,
Living Rev. Relativity {\bf 5}, 1 (2000).

\bibitem{BaumgS03}
T. W. Baumgarte and S. L. Shapiro,
Phys. Rep. {\bf 376}, 41 (2003).

\bibitem{CookP04}
G. B. Cook and H. P. Pfeiffer,
Phys. Rev. D {\bf 70}, 104016 (2004).

\bibitem{CaudiCGP06}
M. Caudill, G. B. Cook, J. D. Grigsby, and H. P. Pfeiffer,
Phys. Rev. D {\bf 74}, 064011 (2006).

\bibitem{GourgGTMB01}
E. Gourgoulhon, P. Grandcl\'ement, K. Taniguchi, J.-A. Marck, and
S. Bonazzola,
Phys. Rev. D {\bf 63}, 064029 (2001).

\bibitem{TanigG02}
K. Taniguchi and E. Gourgoulhon,
Phys. Rev. D {\bf 66}, 104019 (2002).

\bibitem{TanigG03}
K. Taniguchi and E. Gourgoulhon,
Phys. Rev. D {\bf 68}, 124025 (2003).

\bibitem{BejgeGGHTZ05}
M. Bejger, D. Gondek-Rosi\'nska, E. Gourgoulhon, P. Haensel,
K. Taniguchi, and J. L. Zdunik,
Astron. Astrophys. {\bf 431}, 297 (2005).

\bibitem{BonazGM98}
S. Bonazzola, E. Gourgoulhon, and J.-A. Marck,
Phys. Rev. D {\bf 58}, 104020 (1998).

\bibitem{Cook02}
G. B. Cook, Phys. Rev. D {\bf 65}, 084003 (2002).

\bibitem{AshteK04}
A. Ashtekar and B. Krishnan, Living Rev. Relativity {\bf 7}, 10 (2004).

\bibitem{GourgJ06}
E. Gourgoulhon and J. L. Jaramillo,
Phys. Rep. {\bf 423}, 159 (2006).

\bibitem{Lorene}
LORENE web page, http://www.lorene.obspm.fr/.

\bibitem{Blanc02}
L. Blanchet,
Phys. Rev. D {\bf 65}, 124009 (2002);
Living Rev. Relativity {\bf 9}, 4 (2006).

\bibitem{TanigN00}
K. Taniguchi and T. Nakamura,
Phys. Rev. Lett. {\bf 84}, 581 (2000);
Phys. Rev. D {\bf 62}, 044040 (2000).

\bibitem{Zhuge94}
X. Zhuge, J.M. Centrella, and S. L. W. McMillan, Phys. Rev. D {\bf 50},
6247 (1994).

\bibitem{FaberGRT02} 
J. A. Faber, P. Grandcl\'{e}ment, F. A. Rasio, and K. Taniguchi,
Phys. Rev. Lett. {\bf 89}, 231102 (2002).

\bibitem{FaberGR04} 
J. A. Faber, P. Grandcl\'{e}ment, and F. A. Rasio,
Phys. Rev. D {\bf 69}, 124036 (2004).

\bibitem{DenniBP06}
K. A. Dennison, T. W. Baumgarte, and H. P. Pfeiffer, Phys. Rev. D {\bf 74}, 064016 (2006).

\bibitem{angmom}
In \cite{TanigBFS06}, we compared the values of the angular momentum
with the third-order post-Newtonian result, and explained that we found
``better agreement for larger separations''.  In fact, the opposite is correct:
we actually found better agreement at
smaller separations.

\bibitem{GourgGB02}
E. Gourgoulhon, P. Grandcl\'ement, and S. Bonazzola,
Phys. Rev. D {\bf 65}, 044020 (2002);
P. Grandcl\'ement, E. Gourgoulhon, and S. Bonazzola,
Phys. Rev. D {\bf 65}, 044021 (2002).

\end{thebibliography}
\end{document}